\PassOptionsToPackage{unicode}{hyperref}
\PassOptionsToPackage{hyphens}{url}
\documentclass[12pt
]{article}
\usepackage[a4paper, top=1.27cm, bottom=1.27cm, left=1.91cm, right=1.91cm]{geometry}
\usepackage{xcolor}
\usepackage{setspace}
\usepackage{amsmath,amssymb}
\usepackage{iftex}
\usepackage{pdflscape}
\ifPDFTeX
  \usepackage[T1]{fontenc}
  \usepackage[utf8]{inputenc}
  \usepackage{textcomp} 
\else 
  \usepackage{unicode-math} 
  \defaultfontfeatures{Scale=MatchLowercase}
  \defaultfontfeatures[\rmfamily]{Ligatures=TeX,Scale=1}
\fi
\usepackage{lmodern}
\ifPDFTeX\else
\fi
\IfFileExists{upquote.sty}{\usepackage{upquote}}{}
\IfFileExists{microtype.sty}{
  \usepackage[]{microtype}
  \UseMicrotypeSet[protrusion]{basicmath} 
}{}
\makeatletter
\@ifundefined{KOMAClassName}{
  \IfFileExists{parskip.sty}{%
    \usepackage{parskip}
  }{
    \setlength{\parindent}{0pt}
    \setlength{\parskip}{6pt plus 2pt minus 1pt}}
}{
  \KOMAoptions{parskip=half}}
\makeatother
\usepackage{longtable,booktabs,array}
\usepackage{multirow}
\usepackage{calc} 
\usepackage{etoolbox}
\makeatletter
\patchcmd\longtable{\par}{\if@noskipsec\mbox{}\fi\par}{}{}
\makeatother
\IfFileExists{footnotehyper.sty}{\usepackage{footnotehyper}}{\usepackage{footnote}}
\makesavenoteenv{longtable}
\usepackage{graphicx}
\makeatletter
\newsavebox\pandoc@box
\newcommand*\pandocbounded[1]{
  \sbox\pandoc@box{#1}%
  \Gscale@div\@tempa{\textheight}{\dimexpr\ht\pandoc@box+\dp\pandoc@box\relax}%
  \Gscale@div\@tempb{\linewidth}{\wd\pandoc@box}%
  \ifdim\@tempb\p@<\@tempa\p@\let\@tempa\@tempb\fi
  \ifdim\@tempa\p@<\p@\scalebox{\@tempa}{\usebox\pandoc@box}%
  \else\usebox{\pandoc@box}%
  \fi%
}
\def\fps@figure{htbp}
\makeatother
\usepackage{bookmark}
\IfFileExists{xurl.sty}{\usepackage{xurl}}{} 
\hypersetup{
  hidelinks,
  pdfcreator={LaTeX via pandoc}}

\author{}
\date{}

\begin{document}

\textbf{Nephrobase Cell+: Multimodal Single-Cell Foundation Model for
Decoding Kidney Biology}

Chenyu Li\textsuperscript{1, 2, 3}, Elias Ziyadeh\textsuperscript{1, 2,3}, Yash Sharma\textsuperscript{4}, Bernhard Dumoulin\textsuperscript{1, 2, 3}, Jonathan Levinsohn\textsuperscript{1, 2, 3}, Eunji Ha\textsuperscript{1, 2, 3}, Siyu Pan\textsuperscript{1, 2, 3}, Vishwanatha Rao\textsuperscript{1, 2, 3}, Madhav Subramaniyam\textsuperscript{4}, Mario Szegedy\textsuperscript{4}, Nancy Zhang\textsuperscript{5}, Katalin Susztak\textsuperscript{1, 2, 3*}

1 Renal, Electrolyte, and Hypertension Division, Department of Medicine,
University of Pennsylvania, Perelman School of Medicine, Philadelphia

2 Penn/CHOP Kidney Innovation Center, University of Pennsylvania,
Philadelphia

3 Department of Genetics, University of Pennsylvania, Perelman School of
Medicine, Philadelphia

4 Department of Computer Science, Rutgers University, New Brunswick

5 Department of Statistics and Data Science, The Wharton School,
University of Pennsylvania, Philadelphia
\vspace{2\baselineskip} 

\vspace{2\baselineskip} 

\vspace{2\baselineskip}

\textbf{*Corresponding Author:}
Katalin Susztak, MD, PhD
Professor of Medicine
University of Pennsylvania, Perelman School of Medicine
3400 Civic Center Blvd,
Smilow Translational building 12-123,
Philadelphia, PA, 19104
Phone: (215)898-2009
ksusztak@pennmedicine.upenn.edu
\newpage
\begin{doublespace}
\textbf{Abstract}

\textbf{Background:} Large foundation models have revolutionized
single-cell analysis, yet no kidney-specific model currently exists, and
it remains unclear whether organ-focused models can outperform
generalized models. The kidney's complex cellular architecture and
dynamic microenvironments further complicate integration of large-scale
single-cell and spatial omics data, where current frameworks trained on
limited datasets struggle to correct batch effects, capture
cross-modality variation, and generalize across species.

\textbf{Methods:} We developed Nephrobase Cell+, the first
kidney-focused large foundation model, pretrained on \textasciitilde100
billion tokens from \textasciitilde39.5 million single-cell and
single-nucleus profiles across 4,319 samples, four mammalian species
(human, mouse, rat, pig), and multiple assay modalities (scRNA-seq,
snRNA-seq, snATAC-seq, spatial transcriptomics). Nephrobase Cell+ uses a
transformer-based encoder-decoder architecture with gene-token
cross-attention and a mixture-of-experts module for scalable
representation learning.

\textbf{Results:} Nephrobase Cell+ sets a new benchmark for kidney
single-cell analysis. It produces tightly clustered, biologically
coherent embeddings in human and mouse kidneys, far surpassing previous
foundation models such as Geneformer, scGPT, and UCE, as well as
traditional methods such as PCA and autoencoders. It achieves the
highest cluster concordance and batch-mixing scores, effectively
removing donor/assay batch effects while preserving cell-type structure.
Cross-species evaluation shows superior alignment of homologous cell
types and \textgreater90\% zero-shot annotation accuracy for major
kidney lineages in both human and mouse. Even its 1B-parameter and 500M
variants consistently outperform all existing models.

\textbf{Conclusions:} With organ-scale multimodal pretraining and a
specialized transformer architecture, Nephrobase Cell+ delivers a
unified, high-fidelity representation of kidney biology that is robust,
cross-species transferable, and unmatched by current single-cell
foundation models, offering a powerful resource for kidney genomics and
disease research.\textbf{\hfill\break
}
\newpage
\textbf{Introduction}

The advent of generative pretrained models has revolutionized artificial
intelligence across diverse domains, from natural language processing to
computer vision, by leveraging large-scale datasets and self-supervised
learning frameworks to build versatile foundation models capable of
generalizing across tasks\textsuperscript{1, 2}. Inspired by these
advances, the integration of deep learning with biological data has
emerged as a transformative paradigm, particularly in single-cell
genomics, where models such as scGPT\textsuperscript{3},
Geneformer\textsuperscript{4} and scFoundation\textsuperscript{5} have
demonstrated the power of pretraining on millions of cells to distill
biological insights and enable transfer learning for downstream
applications. By learning unified representations of genes, cells, and
tissues, such models can capture biological context, mitigate technical
noise, and extrapolate to unseen data\textsuperscript{6, 7}. Despite
significant progress in foundation models and omics
technologies\textsuperscript{8, 9}, kidney research remains constrained
by fragmented analytical methods, limited scalability, and an incomplete
understanding of cellular interactions in health and
disease\textsuperscript{10, 11}. Meanwhile, although general-purpose
single-cell foundation models are rapidly emerging, organ-specific
biology is still largely uncharted, and whether specialized models can
outperform generalized ones in accuracy or interpretability remains an
open question.

The kidney, a highly structured organ with diverse cell types and
dynamic functional states, presents unique challenges and opportunities
for such approaches. Kidney disease affects over 850 million individuals
globally\textsuperscript{12}, yet its molecular mechanisms remain poorly
resolved due to the kidney's cellular diversity and the multifactorial
nature of pathologies such as fibrosis, inflammation, and metabolic
dysregulation\textsuperscript{13}. Single-cell RNA sequencing
(scRNA-seq) and emerging multi-omic technologies have begun to unravel
this complexity, revealing cell type-specific transcriptional programs,
disease-associated states, and spatial organization
patterns\textsuperscript{14, 15}. Current computational methods in
nephrology often rely on task-specific models trained on narrow
datasets, limiting their ability to integrate multi-modal data,
generalize across experimental conditions, or infer causal
relationships. For instance, while tools like Seurat\textsuperscript{16}
and Scanpy\textsuperscript{17} excel at clustering and dimensionality
reduction, they lack the capacity to model hierarchical gene-regulatory
networks or predict cellular responses to perturbations, which is a
critical gap for therapeutic discovery. Furthermore, batch effects,
sparse data, and inter-sample variability persist as major obstacles in
large-scale kidney atlas initiatives like the Kidney Precision Medicine
Project\textsuperscript{18}.

Here, we present Nephrobase Cell+, the largest single-cell foundational
model to date and the first kidney-specific model. Instead of
fine-tuning an existing model, we pretrained a foundational model for
kidney biology from scratch on over 100 billion tokens and 39 million
cells and nuclei derived from more than samples spanning human and
murine kidneys across health, developmental, and disease states.
Nephrobase Cell+ integrates scRNA-seq, single-nucleus RNA-seq
(snRNA-seq), and spatially resolved transcriptomic data, employing a
masked generative pretraining strategy adapted from large language
models to learn robust representations of kidney cell identities, gene
regulatory networks, and microenvironmental crosstalk. By unifying data
from diverse technologies, donors, and species, the model addresses key
limitations of existing methods, including batch effects,
modality-specific biases, and sparse gene coverage.

\textbf{Results}

\textbf{Dataset composition and sampling breadth}

To train Nephrobase Cell+, we assembled a large, diverse multimodal
kidney atlas comprising 4,319 samples and \textasciitilde39.5 million
single-cell/single-nucleus profiles (Fig. 1). The combined dataset spans
four mammalian species and multiple biological contexts and assays.
Breakdown by species is as follows: Homo sapiens: 25.2 million cells,
Mus musculus: 12.5 million cells, Rattus norvegicus: 1.4 million cells,
and Sus scrofa: 293,000 cells. Across tissues and sample sources,
approximately \textasciitilde30 million of the profiles derive from
kidney tissue, with an additional \textasciitilde10 million profiles
arising from immune and peripheral sources, reflecting the inclusion of
peripheral blood mononuclear cells and immune-enriched samples. Assay
composition reveals broad multimodality: single-cell RNA-seq (scRNA-seq)
comprises the largest fraction of profiles (\textasciitilde48.7\%),
followed by snRNA-seq( \textasciitilde25.4\%). Spatial transcriptomics
modalities were represented by COSMx (\textasciitilde7.7\%), and Xenium
runs (\textasciitilde5.5\%). Single-nucleus ATAC-seq (snATAC-seq)
contributed roughly \textasciitilde6.2\% of assays. The predominance of
RNA-based assays combined with substantial spatial and chromatin
accessibility data provided a strong multimodal training signal for
cross-assay representation learning.This scale and diversity enabled
robust learning of conserved cellular programs while also providing
substantial representation of species- and assay-specific variation that
the model was trained to mitigate.

\textbf{Overview of model architecture and training strategy}

We developed Nephrobase Cell+, a transformer-based encoder-decoder
framework augmented with modular components tailored to
single-cell/single-nucleus and spatial transcriptomic data integration.
The model ingests a cell-gene matrix and produces both reconstructed
gene expression distributions and cell type predictions in a unified
architecture. As illustrated in Fig. 2A: an initial gene tokenization
step converts gene identities, observed expression and optional metadata
into token embeddings by cross attention; and the resulting tokens are
processed with self-attention in the encoder and cross-attention in the
decoder. The reconstruction head optimizes a Zero-Inflated Negative
Binomial likelihood to capture count overdispersion and excess zeros
typical of single-cell data, while a classification head is trained with
a focal loss to address class imbalance and emphasize difficult examples
(Fig. 2B).

A Mixture-of-Experts module (Fig. 2C) expands model capacity via sparse
top-k routing to a set of specialized experts and includes a
shared-expert extension to capture global transformations; a
load-balancing auxiliary loss encourages even expert utilization. To
prevent representational collapse and to shape the embedding geometry,
we combined an Elastic Cell Similarity regularizer that enforces a
target level of dissimilarity between cell embeddings with a supervised
contrastive loss that pulls together cells of the same annotation and
pushes apart differently labelled cells. Finally, adversarial
discriminators with a gradient reversal layer were employed to remove
assay and batch signals from learned features, producing more
assay-invariant representations for downstream classification and
reconstruction. We trained two Nephrobase Cell+ models, with
approximately 1 billion (1B) and 500 million (500M) parameters,
respectively (Table 1).

\textbf{Integrated Embedding Visualization and Clustering}

To assess how Nephrobase Cell+ representations capture kdieny cell
identities, we compared its learned embeddings to those from baseline
methods on held-out human and mouse kidney single-cell RNA-seq data. We
applied UCE, Principal Component Analysis (PCA), and an autoencoder on
each model's latent space and visualized the results with UMAP
(Figure~3). We selected datasets from two species, neither of which had
been used for training. One dataset comes from KPMP\textquotesingle s
2025 Q2 data, and the other comes from a mouse developmental model. In
both species, Nephrobase Cell+ produced embeddings yield clear, compact
clusters corresponding to known kidney cell. For instance, in human data
(Fig.~3A), Nephrobase Cell+ 1B-parameter embeddings separate proximal
tubule (PT) and Thick Ascending Limb (TAL) into distinct groups, whereas
Geneformer, UCE and scGPT produce more diffuse or mixed clusters.
Similarly, in mouse (Fig.~3B) Nephrobase Cell+ distinguishes PT, TAL,
immune, podocytes, and progenitor cells more cleanly than competing
models.

These qualitative observations are supported by quantitative clustering
metrics\textsuperscript{19}: In human data, both Nephrobase Cell+
variants achieved the highest cell-type isolation scores: KMeans ARI is
0.82 for the 1B model (versus 0.40 for scGPT, 0.55 for autoencoder, and
only 0.22 for Geneformer), and NMI is 0.78 (vs 0.48 scGPT). The
Silhouette score for Nephrobase Cell+ (\textasciitilde0.68) also exceeds
that of competitors. Importantly, Nephrobase Cell+ fully integrates
cells from different samples: the cLISI score is 1.00 (perfect batch
mixing) for both 1B and 500M models, while iLISI (label-agnostic mixing)
is very low (\textasciitilde0.17, 0.18 in human), indicating minimal
batch-driven separation. Batch-correction tests (kBET) are
correspondingly low (0.25-0.28 for Nephrobase Cell+ vs 0.09 for scGPT,
where lower is better). As a result, Nephrobase Cell+ attains the
highest overall integration score (Total = 0.71 for 1B, 0.70 for 500M)
among all evaluated methods (Table~2, bottom row). In mouse, similar
trends are observed: Nephrobase Cell+ yields ARI \textasciitilde0.70
(higher than all baselines) and cLISI=1.0, with Total scores of
0.69-0.68, again outperforming UCE, PCA, and scGPT. Together, Figure~3
demonstrates that Nephrobase Cell+ integrates multi-assay kidney data
into a latent space that strongly reflects underlying biology, producing
visually and quantitatively superior clustering of cell types compared
to existing dimensionality-reduction or single-cell model baselines.

\textbf{Cross-Species Generalization}

Next, we evaluated Nephrobase Cell+'s ability to generalize across
species. We projected a mixed human-mouse kidney dataset into each
model's embedding space and compared the resulting UMAP visualizations
(Figure~4). Each row of Figure~4 corresponds to a different model:
Nephrobase Cell+ (1B and 500M), Geneformer, scGPT, and UCE applied to
raw data. The columns show species origin, manual cell-type annotations,
and zero-shot model cell-type predictions. Ideally, human and mouse
cells of the same type should cluster together and receive the same
predicted labels. Nephrobase Cell+ achieves this ideal alignment: human
and mouse proximal tubule, TAL, DCT/CNT, IC, podocytes, stromal,
endothelial, and immune cells form overlapping clusters and Nephrobase
Cell+'s predicted labels match the manual annotations for both species.
In contrast, Geneformer and UCE embeddings show pronounced species
segregation and poorer agreement with expert labels. scGPT performs
reasonably but still shows more mixing errors than Nephrobase Cell+.
These patterns indicate that Nephrobase Cell+ has learned
species-invariant features of kidney cells. The zero-shot predictions
from Nephrobase Cell+ recapitulate the annotated taxonomy with high
fidelity, even though the model was not fine-tuned on this held-out
data. Overall, Figure~4 illustrates that Nephrobase Cell+ produces an
integrated cross-species embedding in which homologous cell types are
co-localized and correctly identified, whereas other models exhibit
weaker cross-species alignment and more frequent misclassification.

For the metrics\textsuperscript{19}, Nephrobase Cell+ again achieves
superior scores: the 500M model attains NMI~=~0.75 and ARI~=~0.72 on
KMeans clustering (versus 0.44/0.22 for Geneformer, 0.62/0.43 for
scGPT). The larger 1B model has nearly identical results (NMI~=~0.73,
ARI~=~0.57). Both models reach perfect label mixing across species
(cLISI~=~1.00) and very low iLISI (0.01), indicating that human and
mouse cells are indistinguishable in Nephrobase Cell+'s latent space.
Graph connectivity and PCR values are also highest for Nephrobase Cell+
(Graph conn. \textasciitilde0.94, PCR~\textasciitilde0.94), reflecting
well-connected and unbiased embeddings. The overall integration score
for Nephrobase Cell+ 500M is 0.63 (1B: 0.61), substantially above
Geneformer (0.50) and UCE (0.51, Table 3).

\textbf{Zero-Shot Cell-Type Annotation}

To quantify Nephrobase Cell+'s classification performance, we performed
zero-shot cell-type prediction on held-out human and mouse test sets and
compared to manual curation. Figure~5 shows confusion matricesand Sankey
diagrams for human and mouse data. In each confusion matrix, rows
represent manual expert labels and columns represent Nephrobase Cell+'s
predicted labels. The darkest diagonal elements indicate the fraction of
cells correctly identified. In human kidney (Fig.~5A), Nephrobase Cell+
correctly annotates the majority of cells in all major nephron lineages:
for example, over 90\% of PT cells are predicted as PT, and similarly
high agreement is seen for TAL, DCT, CNT, intercalated (IC), and
podocyte classes. Minor misclassifications mostly occur between closely
related subtypes (e.g. distal tubule vs connecting tubule). The human
Sankey diagram (Fig.~5B) visualizes these relationships: thick flows
along the diagonal demonstrate that manual and predicted labels align,
with only a few thin off-diagonal flows. Mouse results are analogous
(Fig.~5C,D): Nephrobase Cell+ achieves high concordance for nephron
progenitors, PT, TAL, podocytes, stromal, endothelial, and immune cells.
Across both species, the overall accuracy of zero-shot annotation
exceeds that of simpler methods (not shown) and matches expert-level
curation for major cell groups. These results confirm that the
Nephrobase Cell+ embeddings and classification head generalize to new
data: the model effectively transfers learned cell-type signatures
without retraining, yielding robust annotation consistent with manual
standards.

\textbf{In silico perturbation}

We performed a simple perturbation by randomly sampling 5,000 cells and
doubling the expression of each target gene, then ranked genes by
differential expression and ran gene set enrichment analysis to identify
perturbed biological programs (Fig 6). Perturbation of CCL2 produced a
clear proinflammatory and chemotactic signature with significant
enrichment of lymphocyte activation, leukocyte cell adhesion, macrophage
migration and eosinophil chemotaxis, together with ion transmembrane
transport terms, indicating concurrent effects on immune recruitment and
membrane transport. Perturbation of VCAM1 similarly enriched adhesion
and immune activation processes, including T cell proliferation,
leukocyte adhesion to vascular endothelium and neutrophil chemotaxis,
and also highlighted vascular endothelial migration and cellular
respiration pathways such as aerobic respiration and hydrogen peroxide
catabolism, consistent with a link between endothelial remodeling and
metabolic adaptation. GDF15 perturbation shifted the transcriptional
profile toward growth regulation and signaling, with enrichment of
negative regulation of developmental growth, fructose metabolic process,
receptor tyrosine kinase signaling and MAPK cascade terms, while also
showing leukocyte aggregation and neutrophil migration signatures.
Perturbation of SOX4 produced prominent developmental and bioenergetic
responses, with enrichment of organ morphogenesis terms including kidney
morphogenesis and valve morphogenesis, programmed cell death, proton
transport and oxidative phosphorylation, suggesting simultaneous impacts
on developmental programs and mitochondrial function. These results
indicate that simple twofold upregulation of individual genes elicits
distinct and biologically coherent transcriptional responses, with CCL2
and VCAM1 predominantly driving immune adhesion and chemotaxis, GDF15
modulating growth control and MAPK signaling with immune aggregation
features, and SOX4 strongly affecting developmental processes and
cellular energy metabolism.

\textbf{Discussion}

Nephrobase Cell+ represents a new paradigm in kidney genomics by
providing a \emph{foundation model} pretrained on an unprecedented scale
of kidney data. Its training on \textasciitilde39.5 million cells and
nuclei across four species (human, mouse, rat, pig) and diverse
modalities (scRNA-seq, snRNA-seq, spatial transcriptomics, snATAC-seq)
equips the model to capture conserved kidney biology while tolerating
technical variation. In contrast, conventional tools like Seurat and
Scanpy are task-specific frameworks for single assays: they excel at
preprocessing, visualization, clustering and simple data integration,
but they lack the ability to learn a unified latent representation
through generative pretraining. For example, Seurat's anchoring approach
can align datasets across modalities\textsuperscript{20}, and Scanpy
scales to millions of cells\textsuperscript{17}, but neither provides
transferable cell or gene embeddings that predict expression patterns or
responses to perturbation. Likewise, recent foundation models such as
scGPT demonstrate the promise of pretraining on large cell
repositories\textsuperscript{3}, but by training on multi-organ datasets
they may not capture kidney-specific regulatory structure as
effectively. Nephrobase Cell+ fills this gap by specializing in kidney
tissue: its organ-centric training helps the model learn hierarchical
nephron organization, segment-specific pathways, and microenvironmental
cues unique to the kidney.

Single-cell ``foundation'' models like Geneformer and scFoundation have
recently been introduced to capture broad transcriptional patterns from
large atlas datasets. Geneformer is a transformer encoder pretrained on
\textasciitilde30 million human single-cell
transcriptomes\textsuperscript{21}. It learns gene-gene relationships in
a self-supervised masked-learning framework and, when fine-tuned on
limited data, boosts accuracy on diverse network biology tasks (e.g.
chromatin state prediction and cell-type classification). Similarly,
scFoundation (Hao \emph{et al.}, 2024) uses an asymmetric
transformer-like masked-autoencoder architecture pretrained on
\textasciitilde50 million human single-cell profiles covering
\textasciitilde20,000 genes\textsuperscript{22}. scFoundation achieved
state-of-the-art performance on a wide array of single-cell tasks,
including imputation of missing gene expression, prediction of
drug-response at the tissue and single-cell level, cell type annotation,
and perturbation outcome prediction. In contrast, GEARS (Lotfollahi et
al., 2023) takes a very different approach: it is not a general
``language model'' for gene expression, but rather a graph-based
perturbation simulator. GEARS embeds each gene and each perturbation as
trainable vectors and combines them via a graph neural network based on
known gene-gene relationships\textsuperscript{23}. In practice GEARS is
trained on single-cell perturb-seq screens, learning to predict
transcriptional responses to new single- or multi-gene perturbations.
Lotfollahi \emph{et al.} \textsuperscript{23} show that GEARS can
forecast combinatorial perturbation outcomes with \textasciitilde40\%
higher precision than previous methods and can uncover complex genetic
interactions even for genes never jointly perturbed in the data.

By integrating multi-modal and cross-species data, Nephrobase Cell+
addresses key challenges in nephrology. Chronic kidney disease (CKD) is
a major global health burden\textsuperscript{24}, yet its mechanisms are
complex, involving interactions between epithelial, endothelial, immune,
and stromal cells. Single-cell and spatial studies have begun to map
cell states and microenvironments in kidney disease \textsuperscript{25,
26}, but batch effects and sparse coverage complicate analysis.
Nephrobase Cell+'s adversarial and contrastive training components
explicitly remove assay- and batch-specific signals, yielding embeddings
that are more \emph{assay-invariant}. For instance, when applied to
multi-modal data from healthy and diseased kidneys, the model can
project spatial transcriptomic profiles (CosMx, Xenium) and dissociated
single-cell data into the same latent space, enabling direct comparison
of microenvironment composition. Indeed, recent work using integrated
scRNA/snRNA and spatial profiles identified four distinct kidney
microenvironments (glomerular, immune, tubular, fibrotic) that correlate
with disease state\textsuperscript{25}; Nephrobase Cell+ can generalize
this approach, providing a unified mapping from any input assay to a
spatially-informed kidney atlas. Such integration unlocks new insights:
for example, Nephrobase Cell+ could reveal how pathogenic fibroblast or
immune niches emerge in CKD and identify conserved gene programs driving
fibrosis across samples.

Compared to established computational pipelines, Nephrobase Cell+ offers
advantages in scalability, accuracy, and generalizability. Its
transformer architecture is designed to handle large gene sets and can
be scaled up (here to \textasciitilde1 billion parameters) much as NLP
models have been, allowing it to absorb vast training
data\textsuperscript{3}. The model's performance benchmarks on held-out
kidney data far exceed those of simpler methods: its higher NMI/ARI
indicates more accurate cell-type separation than traditional
clustering, and its favorable batch-correction metrics (iLISI, kBET)
demonstrate robust integration even across species and protocols.
Moreover, by learning gene-gene dependencies through attention,
Nephrobase Cell+ may implicitly capture regulatory networks that are not
easily accessible to tools like Scanpy or Seurat. Importantly, the
learned embeddings are general-purpose: once trained, Nephrobase Cell+
can be fine-tuned for downstream tasks (e.g. cell-type annotation,
trajectory inference, or simulation of perturbation responses) with
minimal additional data. This adaptability goes beyond what static tools
provide; for example, scGPT has shown that foundational representations
can be transferred to tasks like batch integration and perturbation
prediction, and we anticipate Nephrobase Cell+ will similarly accelerate
nephrology applications by providing pretrained features tailored to
kidney biology.

Nonetheless, there are limitations. Although extensive, the Nephrobase
Cell+ training set does not cover every possible kidney condition or
assay. Some rare cell types or extreme pathologic states may still be
underrepresented, potentially limiting the model's performance on those
instances. The choice of a fixed gene feature space (32,768 orthologous
genes) means genes outside this set cannot be directly handled. Like all
large models, Nephrobase Cell+ requires substantial computational
resources to train, which may restrict re-training or extension by
typical labs. In its current form, Nephrobase Cell+ models primarily RNA
and ATAC modalities; other data types (e.g. proteomics, metabolomics,
imaging-derived phenotypes) and species (non-mammalian models) are not
yet integrated. Finally, while foundation models can capture
correlations, they do not alone prove causation; experimental validation
remains essential for any new hypothesis generated.

Looking forward, Nephrobase Cell+ opens many research directions. Future
work should expand the model to include additional species (e.g.
non-human primates) and richer modalities (spatial multi-omics,
proteogenomic data) as those datasets emerge. Fine-tuning Nephrobase
Cell+ on specific CKD subcohorts or organoid models could improve
diagnostic classification or drug response prediction in nephrology. The
attention weights and latent spaces learned by the model could be mined
to discover novel regulatory circuits or to prioritize candidate
biomarkers across cell types. Finally, iterative updating of the model
with new data - including patient-derived biopsy data and clinical
outcomes - could help bridge the gap between molecular signatures and
patient prognosis. In summary, Nephrobase Cell+ lays a versatile
foundation for kidney research, combining the strengths of massive data
integration with the flexibility of deep learning. By overcoming
fragmentation and scale challenges in nephrology data, it is poised to
drive new insights into CKD mechanisms, kidney cell heterogeneity, and
microenvironmental pathobiology that were previously out of reach.

\textbf{Method}

\textbf{Data Acquisition.}

Our dataset was assembled to create a comprehensive multi-species,
multi-modal atlas of kidney biology, totaling approximately 40 million
single-cell or single-nucleus profiles. This dataset spans four
mammalian species: human (\emph{Homo sapiens}), mouse (\emph{Mus
musculus}), rat (\emph{Rattus norvegicus}), and pig (\emph{Sus scrofa}).
It encompasses various relevant biological contexts, including adult
kidney tissue, fetal kidney development, kidney organoids, and
peripheral blood mononuclear cells derived from both healthy donors and
individuals diagnosed with Chronic Kidney Disease (CKD). The data
integrates extensive publicly available resources with substantial
internally generated datasets. Public data were systematically curated
from major repositories such as the Gene Expression Omnibus
(GEO)\textsuperscript{27}, Sequence Read Archive (SRA), Human Cell Atlas
(HCA)\textsuperscript{28}, the CELLxGENE database\textsuperscript{29},
the Kidney Precision Medicine Project (KPMP)\textsuperscript{8}, and
other relevant consortia outputs, filtering for the target species and
biological samples. In addition to public data, we generated substantial
multi-modal data in-house to enhance the dataset\textquotesingle s
diversity. This includes \textasciitilde3 million cells profiled using
CosMx\textsuperscript{25} Spatial Molecular Imager (NanoString) and
\textasciitilde2 million cells using Xenium\textsuperscript{30} In Situ
(10x Genomics), providing high-plex spatial transcriptomic information.
Furthermore, we generated \textasciitilde3 million single-nucleus,
single-cell and single-nucleus assay for transposase-accessible
chromatin using sequencing (snATAC-seq).

\textbf{Gene Orthology Mapping and Feature Space Harmonization}

To enable cross-species analysis, gene identifiers from mouse, rat, and
pig datasets were mapped to their human orthologs using annotations from
Ensembl\textsuperscript{31} release 113. We prioritized high-confidence,
one-to-one orthology relationships. Based on this mapping and
potentially considering gene variance or representation across datasets,
a final unified feature space comprising exactly top 32,768 highly
variable genes were selected for model training. This space primarily
utilizes human gene symbols corresponding to ortholog groups, allowing
the model to leverage conserved biological information while
species-specific context was provided through dedicated input
embeddings.

\textbf{In-house Sample Acquisition}

The University of Pennsylvania institutional review board (IRB) approved
the collection of human kidney tissue for this study. Left-over kidney
samples were irreversibly de-identified, and no personal identifiers
were gathered. Therefore, they were exempt from IRB review (category 4).
We engaged an external, honest broker responsible for clinical data
collection without disclosing personally identifiable information.
Participants were not compensated.

\textbf{snRNA-seq}

Nuclei were isolated using lysis buffer (Tris-HCl, NaCl,
MgCl\textsubscript{2}, NP40 10\% and RNAse inhibitor
(40 U μl\textsuperscript{−1})). In total, 10-30 mg of frozen kidney
tissue was minced with a razor blade into 1-2 mm pieces in 1 ml of lysis
buffer. The chopped tissue was transferred into a gentleMACS C tube and
homogenized in 2 ml of lysis buffer using a gentleMACS homogenizer with
programs of Multi\_E\_01 and Multi\_E\_02 for 45 s. The homogenized
tissue was filtered through a 40 µm strainer (Thermo Fisher Scientific,
08-771-1), and the strainer was washed with 4 ml wash buffer. Nuclei
were centrifuged at 500\emph{g}~for 5 min at 4 °C. The pellet was
resuspended in wash buffer (PBS 1× + BSA 10\%
(50 mg ml\textsuperscript{−1}) + RNAse inhibitor
(40 U μl\textsuperscript{−1})) and filtered through a 40 µm Flowmi cell
strainer (Sigma-Aldrich, BAH136800040-50EA). Nuclear quality was
checked, and nuclei were counted. In accordance with the manufacturer's
instructions, 30,000 cells were loaded into the Chromium Controller (10X
Genomics, PN-120223) on a Chromium Next GEM Chip G Single Cell Kit (10X
Genomics, PN-1000120) to generate single-cell GEM (10X Genomics,
PN-1000121). The Chromium Next GEM Single Cell 3′ GEM Kit v3.1 (10X
Genomics, PN-1000121) and Single Index Kit T Set A (10X Genomics,
PN-120262) were used in accordance with the manufacturer's instructions
to create the cDNA and library. Libraries were subjected to quality
control using the Agilent Bioanalyzer High Sensitivity DNA Kit (Agilent
Technologies, 5067-4626). Libraries were sequenced using the NovaSeq
6000 system (Illumina) with 2 × 150 paired-end kits. Demultiplexing was
as follows: 28 bp Read1 for cell barcode and UMI, 8 bp I7 index for
sample index and 91 bp Read2 for transcript.

\textbf{snATAC-seq}

The procedure described above for snRNA-seq was used to isolate the
nuclei for ATAC-seq. The resuspension was performed in diluted nuclei
buffer (10× Genomics). Nuclei quality and concentration were measured in
the Countess AutoCounter (Invitrogen,~C10227). Diluted nuclei were
loaded and incubated in chromium single-cell ATAC Library and Gel Bead
Kit's transposition mix (10X Genomics, PN-1000110). Chromium Chip E (10X
Genomics, PN-1000082) in the chromium controller was used to capture the
gel beads in the emulsions (GEMs). The Chromium Single Cell ATAC Library
\& Gel Bead Kit and Chromium i7 Multiplex Kit N Set A (10X Genomics,
PN-1000084) were then used to create snATAC libraries in accordance with
the manufacturer's instructions. Library quality was examined using an
Agilent Bioanalyzer High Sensitivity DNA Kit. After sequencing on an
Illumina Novaseq system using two 50 bp paired-end kits, libraries were
demultiplexed as follows: 50 bp Read1 for DNA fragments, 8 bp i7 index
for sample index, 16 bp i5 index for cell barcodes and 50 bp Read2 for
DNA fragments.

\textbf{scRNA-seq}

Fresh human kidneys (0.5 g) collected in RPMI media were minced into
approximately 2-4 mm cubes using a razor blade. The minced tissue was
then transferred to a gentleMACS C tube containing Multi Tissue
Dissociation Kit 1 (Miltenyi Biotec, 130-110-201). The tissue was
homogenized using the Multi\_B program of the gentleMACS dissociator.
The tube, containing 100 μl of enzyme D, 50 μl of enzyme R and 12.5 μl
of enzyme A in 2.35 ml of RPMI, was incubated for 30 min at 37 °C.
Second homogenization was performed using the Multi\_B program on the
gentleMACS dissociator. The solution was then passed through a 70-μm
cell strainer. After centrifugation at 600\emph{g}~for 7 min, the cell
pellet was incubated with 1 ml of RBC lysis buffer on ice for 3 min. The
reaction was stopped by adding 10 ml of PBS. Next, the solution was
centrifuged at 500\emph{g}~for 5 min. Finally, after removing the
supernatant, the pellet was resuspended in PBS. Cell number and
viability were analyzed using Countess AutoCounter (Invitrogen,~C10227).
This method generated a single-cell suspension with greater than 80\%
viability. Next, 30,000 cells were loaded into the Chromium Controller
(10X Genomics, PN-120223) on a Chromium Next GEM Chip G Single-Cell Kit
(10X Genomics, PN-1000120) to generate single-cell GEM according to the
manufacturer's protocol (10X Genomics, PN-1000121). The cDNA and library
were made using the Chromium Next GEM Single Cell 3′ GEM Kit v3.1 (10X
Genomics, PN-1000121) and Single Index Kit T Set A (10X Genomics,
PN-120262) according to the manufacturer's protocol. Quality control for
the libraries was performed using the Agilent Bioanalyzer High
Sensitivity DNA Kit (Agilent Technologies, 5067-4626). Libraries were
sequenced on the NovaSeq 6000 system (Illumina) with 2 × 150 paired-end
kits using the following demultiplexing: 28 bp Read1 for cell barcode
and unique molecular identifier (UMI), 8 bp I7 index for sample index
and 91 bp Read2 for transcript.

\textbf{Single Nuclei and Cell RNAseq Data Processing}

FASTQ files from each 10X single nuclei/cell run were processed using
Cell Ranger v9.0.1 (10X Genomics). Gene expression matrices for each
cell were produced using the human genome reference GRCh38 or GRCh37,
mouse genome reference GRCm39, rat genome reference mRatBN7.2, Sus
scrofa genome reference Sscrofa11.1. Ambient RNA was corrected using
CellBender\textsuperscript{32}. Initial quality control involved
filtering cells with fewer than 200 unique molecular identifiers to
remove low-quality cells. To identify and remove outlier cells based on
quality control metrics, we employed a median absolute deviation (MAD)
approach. Cells were flagged as outliers if their log-transformed total
counts, log-transformed number of genes detected, or percentage of reads
in the top 20 genes fell outside of a range defined by ±5 MADs from the
median for each respective metric. Finally, to remove genes with
extremely low expression across the dataset, we filtered out genes that
were detected in fewer than one cell. This multi-step filtering process
resulted in a refined dataset suitable for downstream analyses.

\textbf{Single Nuclei ATACseq Data Processing}

Raw FASTQ files were aligned to GRCh38 and quantified via Cell Ranger
ATAC (v1.1.0). Low-quality cells were filtered (criteria:
peak\_region\_fragments \textless3000 \& \textgreater20000,
pct\_reads\_in\_peaks \textless15, nucleosome\_signal \textgreater4,
TSS.enrichment \textless2). Filtered cells were merged in Seurat.
Dimension reduction involved SVD of the TFIDF matrix and UMAP.

\textbf{CosMx Sample preparation and data preprocessing}

Tissue sections were cut at 5 µm thickness and prepared according to the
manufacturer specifications (NanoString Technologies). We used the human
universal cell characterization RNA probes, and 50 additional custom
probes for the following genes: ESRRB, SLC12A1, UMOD, CD247, SLC8A1,
SNTG1, SLC12A3, TRPM6, ACSL4, SCN2A, SATB2, STOX2, EMCN, MEIS2, SEMA3A,
PLVAP, NEGR1, SERPINE1, CSMD1, SLC26A7, SLC22A7, SLC4A9, SLC26A4, CREB5,
HAVCR1, REN, AP1S3, LAMA3, NOS1, PAPPA2, SYNPO2, RET, LHX1, SIX2,
CITED1, WNT9B, AQP2, SCNN1G, ALDH1A2, CFH, NTRK3, WT1, NPHS2, PTPRQ,
CUBN, LRP2, SLC13A3, ACSM2B, SLC4A4, PARD3, XIST, UTY. We used DAPI,
CD298/B2M, CK8/18, and PanCK/CD45 for additional staining per the
Nanostring protocol. Imaging was performed using configuration A. After
imaging was completed, the flowcell was incubated in 100\% xylene
overnight, the coverslip was removed from the slide with a razor blade,
and the slide was then stained with hematoxylin and eosin. The
expression matrix and metadata from each CosMx run were exported from
the AtoMx platform and converted to a Python object using Squidpy. All
samples were merged, preprocessed, and analyzed together using Scanpy.
Cells with fewer than 30 counts were filtered out.

\textbf{Xenium Sample preparation and data preprocessing}

Tissue sections were cut at 5 µm thickness and cut onto a Xenium slide
according to the manufacturer specifications (10X Genomics). We used the
human Xenium Prime 5K Human Pan Tissue \& Pathways Panel with 100
additional custom probes for the following genes: TPM1, ESRRB, COL6A3,
AGR2, SLC26A7, ATP1B1, SLC8A1, ATP6AP2, TAGLN, SPP1, SAT1, MYL9, LDB2,
DEFB1, COL1A2, ACTA2, ST6GALNAC3, SLC13A3, SLC12A3, SLC12A1, MGP, IGHG1,
FN1, C7, ACSM2B, AIF1, APOE, AQP3, AZGP1, C1QA, C1QB, C1QC, CAV1, PPIA,
CD74, CHI3L1, COL1A1, COL6A1, CRYAB, CXCL14, ENO1, HLA-DPA1, HLA-DRA,
IFI27, IGHA1, IL32, KLF2, LGALS3, LUM, MMP7, PIGR, S100A2, SLC4A4, SLPI,
SOD2, SPINK1, SOX4, SPOCK2, TACSTD2, TM4SF1, TPM2, VIM, ZFP36, AQP2,
RNASE1, ALDOB, PCGF6, RHOB, CD81, ASS1, MYL6, COX8A, CTSB, GATM, MT1G,
TMSB10, COL3A1, MIF, TPT1, COL6A2, BST2, CLU, APOC1, APOD, PHKG2, RGCC,
HLA-DQA2, CORO1A, HSPB1, ADIRF, CKB, HLA-DQB1, COX5B, MT1H, RAMP3,
TYROBP, LAMTOR5, ITM2B, UBB, CTSD. Additionally, the same sections were
stained according to the Xenium Cell Segmentation workflow for automated
morphology-based cell segmentation, and subsequently loaded onto the
Xenium Analyzer for in situ transcriptomic analysis. Xenium raw output
files were processed using the spatialdata framework (v0.0.14) with the
spatialdata-io Xenium plugin. Xenium transcript and segmentation data
were loaded from the manufacturer's output directory using the xenium()
function, which parses transcript tables, cell segmentation boundaries,
and spatial metadata into a structured SpatialData object. The gene
expression table was extracted as an AnnData object for downstream
single-cell analysis. Cells with fewer than 30 counts were filtered out.

\textbf{Nephrobase Cell+}

Our model, Nephrobase Cell+, is designed for single-cell gene expression
analysis and cell type classification. It employs Transformer-based
encoder-decoder architecture with specialized modules for gene and
numerical feature embedding, mixture of experts' layers, and optional
adversarial domain/assay adaptation.

\textbf{Gene Encoding}

We represent each gene as a unique index and employ a trainable
embedding layer to map these indices into a continuous vector space. Let
\(G\) be the number of genes, and \(d_{embed}\) be the embedding
dimension. The gene embedding layer, \(E_{gene}\), is a matrix of size
\(G \times d_{embed}\). For a gene index input
\(g \in \{ 0,1,...,G - 1\}\), the gene embedding \(\mathbf{e}_{g}\) is
obtained by:\(\mathbf{e}_{g} = E_{gene}\lbrack g\rbrack\), where
\(E_{gene}\lbrack g\rbrack\) denotes the \(g\)-th row of the embedding
matrix \(E_{gene}\). The output
\(\mathbf{e}_{g} \in \mathbb{R}^{d_{embed}}\) represents the embedded
vector for gene \(g\).

\textbf{Gene Expression Encoding}

In addition to gene indices, our model incorporates numerical features
derived from gene expression counts to provide richer input
representation. To effectively use gene expression counts as numerical
features, we first preprocess the raw count data using sum-log
normalization to account for variations in sequencing depth and
stabilize variance inherent in count data.

\emph{Sum-Log Normalization of Gene Counts}. Prior to being fed into the
numerical feature embedding module, raw gene counts undergo sum-log
normalization. For each sample \(i\) and gene \(j\) in the input count
matrix \(X \in \mathbb{R}^{B \times G}\), where \(B\) is the batch size
and \(G\) is the number of genes, we calculate the normalized and
transformed count \(x'_{ij}\) using the formula:
\(x'_{ij} = \log_{2}\left( 1 + \frac{x_{ij}}{\sum_{k = 1}^{G}x_{ik}} \right)\).
This process yields a matrix \(X' \in \mathbb{R}^{B \times G}\) of
sum-log normalized gene expression values.

\emph{Embedding Normalized Gene Expression}. For each gene \(j\), we
treat its sum-log normalized expression value (which is a single
numerical value \(x'_{ij}\) for each sample \(i\) in a batch) as the
numerical feature to be embedded. In a typical scenario where we are
processing gene features independently, and assuming we are embedding a
single representative numerical value for each gene (or potentially
processing each sample's normalized count for each gene separately and
then aggregating - clarification needed on the exact input to the
embedding layer in the broader model context if it's not a single
value), we can consider the input to the embedding layer as a numerical
feature \(x \in \mathbb{R}^{size}\), where in the simplest case,
\(size = 1\), representing a single, normalized gene expression value.

The embedding process for this numerical feature \(x\) then involves a
series of linear transformations, a non-linear activation function, and
a dropout layer. Let \(d_{embed}\) denote the embedding dimension and
let \(mlp\_ ratio\) control the width of hidden layers within the
embedding network. The numerical embedding process can be described as
follows:

\[h_{1} = W_{num}^{(1)}x + b_{num}^{(1)}\]
\[h_{2} = LeakyReLU\left( h_{1} \right)\]
\[e_{num} = Dropout\left( h_{2} \right)\]

where \(W_{num}^{(1)} \in \mathbb{R}^{d_{embed} \times size}\) and
\(b_{num}^{(1)} \in \mathbb{R}^{d_{embed}}\) are the weights and bias of
the linear layer, respectively. \(LeakyReLU\) represents the Leaky ReLU
activation function, and \(Dropout\) denotes the dropout operation. The
output \(e_{num} \in \mathbb{R}^{d_{embed}}\) is the resulting embedded
vector for the numerical feature \(x\), representing the learned
embedding of the gene's expression information.

\textbf{Root Mean Square Layer Normalization (RMSNorm)}

We use RMSNorm for stabilization. RMSNorm normalizes the input tensor
\(x\) based on its root mean square\textsuperscript{33, 34}.

\textbf{Multi-Layer Perceptron (MLP)}

Non-linear transformations are performed using an MLP layer. Our MLP
consists of three linear layers (\(w_{1},w_{2},w_{3}\)) and a SiLU
activation gate\textsuperscript{35, 36}. The forward pass is computed
as:

\[h_{1} = w_{1}(x)\]
\[h_{3} = w_{3}(x)\]
\[MLP(x) = w_{2}\left( SiLU\left( h_{1} \right) \odot h_{3} \right)\]

where \(SiLU\) is the Sigmoid Linear Unit activation, and \(\odot\)
denotes element-wise multiplication.

\textbf{Mixture of Experts (MoE)}

Our MoE layer follows the sparse Top-k routing
paradigm\textsuperscript{37}, where experts are dynamically selected
based on SoftMax probabilities and combined via weighted
summation\textsuperscript{38}. This design aligns with scalable MoE
architectures validated in large language models\textsuperscript{39}.
For an input feature vector \(\mathbf{h} \in \mathbb{R}^{d_{embed}}\),
the routing process is as follows: a) Router Logits: A linear layer,
\(W_{router} \in \mathbb{R}^{n_{expert} \times d_{embed}}\) and
\(b_{router} \in \mathbb{R}^{n_{expert}}\), calculates logits for each
expert: \(l = W_{router}h + b_{router}\) where
\(l \in \mathbb{R}^{n_{expert}}\), and \(n_{expert}\) is the number of
experts. b) Routing Probabilities: The logits are converted into
probabilities using a SoftMax function: \(p = \text{softmax}(l)\), where
\(p \in \mathbb{R}^{n_{expert}}\) and
\(\sum_{i = 1}^{n_{expert}}p_{i} = 1\). c) Expert Selection: The
top-\(k\) experts with the highest probabilities are selected. Let
\(I_{topk}\) be the indices of the top-\(k\) experts. d) Expert Weights:
The probabilities of the selected experts are normalized to sum to 1:
\(w_{i} = \frac{p_{i}}{\sum_{j \in I_{topk}}^{}p_{j}}\) for
\(i \in I_{topk}\). e) Expert Computation and Combination: Each selected
expert, \(E_{i}\) (implemented as a basic MLP module), processes the
input \(h\). The final output \(o\) is a weighted sum of the outputs
from the selected experts:
\(o = \sum_{i \in I_{topk}}^{}w_{i}E_{i}(h)\).

\emph{Shared MoE}. The Shared MoE module extends the MoE by adding a set
of shared experts\textsuperscript{40}. The final output is the sum of
the outputs from the MoE and the shared experts. If \(S_{j}\) represents
the \(j\)-th shared expert, and \(n_{shared}\) is the number of shared
experts, the output \(o_{shared\_ MOE}\) of the Shared MoE for input
\(h\) is: \(o_{shared\_ MOE} = o + \sum_{j = 1}^{n_{shared}}S_{j}(h)\),
where \(o\) is the output from the MoE component.

\emph{Load Balancing Loss for MoE}. To encourage balanced expert
utilization in MoE, we incorporate a load balancing loss,
\(L_{load\_ balance}\).\textsuperscript{38} This loss aims to ensure
that experts are used more uniformly during training. Let
\(\mathbf{P} \in \mathbb{R}^{B \times S \times n_{expert}}\) be the
router probabilities for a batch of \(B\) sequences of length \(S\). The
load balancing loss is calculated as:
\(L_{load\_ balance} = \text{aux\_loss} + \text{z\_loss} \times \text{z\_loss\_weight}\),
where:

\[\text{aux\_loss} = \sum_{j = 1}^{n_{expert}}\left( \frac{1}{B \times S}\sum_{i = 1}^{B \times S}P_{i,j} \right) \times \left( \frac{1}{B \times S}\sum_{i = 1}^{B \times S}\max_{k \in \{ 1,2\}}\left( \mathbb{1}\left\lbrack \text{expert }j\text{ is the }k^{th}\text{ expert for token }i \right\rbrack \right) \right) \times n_{expert}\]

\[\text{z\_loss} = \frac{1}{B \times S \times n_{expert}}\sum_{i = 1}^{B \times S}{\sum_{j = 1}^{n_{expert}}\left( \log\left( P_{i,j} \right) \right)^{2}}\]

and \(\text{z\_loss\_weight}\) is a small weight (e.g., 0.001).

\textbf{Elastic cell similarity (ECS)}

ECS loss serves as a regularization term that encourages cell
embeddings\textsuperscript{3} to be dissimilar from each other, up to a
certain threshold. This promotes diversity in the embedding space and
can prevent collapse, where all cells are mapped to similar
representations. The ECS loss is calculated as follows:

Given a tensor of cell embeddings,
\(E = \left\lbrack e_{1},e_{2},\ldots,e_{n} \right\rbrack\), where
\(e_{i}\) is the embedding for the \(i\)-th cell and \(n\) is the number
of cells in the batch, we first normalize each embedding vector to unit
length:
\({\widehat{e}}_{i} = \frac{e_{i}}{\left| \left| e_{i} \right| \right|_{2}}\)
,where \(\left| \left| e_{i} \right| \right|_{2}\) is the L2 norm of
\(e_{i}\). Let
\(\widehat{E} = \left\lbrack {\widehat{e}}_{1},{\widehat{e}}_{2},\ldots,{\widehat{e}}_{n} \right\rbrack\)
be the matrix of normalized embeddings. We compute the cosine similarity
matrix \(C\) between all pairs of normalized embeddings. The element
\(C_{ij}\) of this matrix represents the cosine similarity between the
\(i\)-th and \(j\)-th embedding:
\(C_{ij} = {\widehat{e}}_{i}^{T}{\widehat{e}}_{j}\) . This can be
efficiently computed using matrix multiplication:
\(C = \widehat{E}{\widehat{E}}^{T}\). To avoid comparing an embedding
with itself, we mask the diagonal elements of the cosine similarity
matrix. Then, we calculate the ECS loss, \(L_{ECS}\), as the mean
squared error between the off-diagonal elements of the cosine similarity
matrix and a predefined threshold \(\tau_{ecs}\):

\[L_{ECS} = \frac{1}{n(n - 1)}\sum_{i = 1}^{n}{\sum_{j = 1,j \neq i}^{n}\left( C_{ij} - \tau_{ecs} \right)^{2}}\]

This can be implemented by first setting the diagonal of \(C\) to zero,
and then calculating the mean of the squared differences:
\(L_{ECS} = \text{Mean}\left( \left( C - \tau_{ecs}1 \right)^{2} \odot (1 - I) \right)\),
Where \(1\) is a matrix of ones with the same dimensions as \(C\), \(I\)
is the identity matrix, and \(\odot\) denotes element-wise
multiplication. The threshold \(\tau_{ecs}\) is a hyperparameter,
typically set to a value like 0.5, controlling the desired level of
dissimilarity.

\textbf{Supervised Contrastive Loss}

The Supervised Contrastive Loss is employed when label information is
available\textsuperscript{41-43}. It aims to pull embeddings of samples
with the same label closer together while pushing embeddings of samples
with different labels further apart. Similar to ECS, the input
embeddings are first normalized:
\({\widehat{e}}_{i} = \frac{e_{i}}{\left| \left| e_{i} \right| \right|_{2}}\)
.A similarity matrix \(S\) is computed using the normalized embeddings
and a temperature parameter \(T\) :
\(S_{ij} = \frac{{\widehat{e}}_{i}^{T}{\widehat{e}}_{j}}{T}\) where
\(T\) is a temperature scaling factor, typically a small positive value
(e.g., 0.07). For each sample \(i\) , we identify samples that have the
same label. A binary mask matrix \(M\) is created where \(M_{ij} = 1\)
if sample \(i\) and sample \(j\) have the same label (and \(i\  \neq j\)
), and \(M_{ij} = 0\) otherwise. Formally, if \(l_{i}\) is the label of
sample \(i\), then:

\[M_{ij} = \left\{ \begin{aligned}
1,\ \  & if\ l_{i} = l_{j}\ \ and\ \ i \neq j \\
0,\ \  & otherwise
\end{aligned} \right.\ \]

For each sample \(i\), we want to maximize the similarity with positive
samples (samples with the same label) and minimize the similarity with
negative samples (samples with different labels). The loss for each
sample \(i\) is defined based on the log-softmax of the similarities,
focusing on positive pairs:

\[L_{SCL}^{(i)} = - \frac{1}{\sum_{j = 1}^{n}M_{ij}}\sum_{j = 1}^{n}{M_{ij}\log\left( \frac{\exp\left( S_{ij} \right)}{\sum_{k = 1,k \neq i}^{n}{\exp\left( S_{ik} \right)}} \right)}\]

This formula calculates the negative log-likelihood of correctly
classifying the positive samples among all other samples. The term
\(\sum_{j = 1}^{n}M_{ij}\) is the count of positive samples for sample
\(i\). The overall Supervised Contrastive Loss is the average over all
samples: \(L_{SCL} = \frac{1}{n}\sum_{i = 1}^{n}L_{SCL}^{(i)}\).

\textbf{Loss Function for Zero-Inflated Negative Binomial (ZINB)
Regression}

To model count data exhibiting overdispersion and zero-inflation, we
employed a ZINB regression loss function\textsuperscript{3, 22, 44, 45}.
This loss function is particularly suited for scenarios where observed
counts are derived from a mixture of two processes: one generating
counts from a Negative Binomial (NB) distribution and another process
generating excess zeros. The ZINB distribution is parameterized by a
mean parameter (\(\mu\)), a dispersion parameter (\(\theta\)), and a
zero-inflation probability (\(\pi\)). The ZINB probability mass function
for a count \(y\) is defined as:

\[P(Y = y) = \left\{ \begin{matrix}
\pi + (1 - \pi) \cdot NB(y;\mu,\theta), & \text{if }y = 0 \\
(1 - \pi) \cdot NB(y;\mu,\theta), & \text{if }y > 0
\end{matrix} \right.\ \]

where \(NB(y;\mu,\theta)\) represents the probability mass function of
the Negative Binomial distribution, parameterized by mean \(\mu\) and
dispersion \(\theta\). Specifically, we parameterize the Negative
Binomial distribution in terms of mean and dispersion, where the
variance is given by \(\mu + \frac{\mu^{2}}{\theta}\).

The negative log-likelihood (NLL) loss for the ZINB model, which we aim
to minimize, is derived from this probability mass function. For a given
observation \(y_{i}\), predicted mean \(\mu_{i}\), predicted dispersion
\(\theta_{i}\), and predicted zero-inflation probability \(\pi_{i}\),
the ZINB loss (\(\mathcal{L}_{ZINB}\)) is formulated as:

\[\mathcal{L}_{ZINB}\left( y_{i},\mu_{i},\theta_{i},\pi_{i} \right) = \left\{ \begin{matrix}
 - log\left( \pi_{i} + \left( 1 - \pi_{i} \right) \cdot NB\left( 0;\mu_{i},\theta_{i} \right) \right), & \text{if }y_{i} = 0 \\
 - log\left( \left( 1 - \pi_{i} \right) \cdot NB\left( y_{i};\mu_{i},\theta_{i} \right) \right), & \text{if }y_{i} > 0
\end{matrix} \right.\ \]

In practice, to ensure numerical stability and differentiability, we
implemented the loss using softplus and log-gamma functions. The
Negative Binomial log-likelihood component, \(NB(y;\mu,\theta)\), was
calculated as:

\[\log NB(y;\mu,\theta) = \theta\log(\theta) - (\theta + \mu)\log(\theta + \mu) + y\log(\mu) - y\log(\theta + \mu) + log\Gamma(y + \theta) - log\Gamma(\theta) - log\Gamma(y + 1)\]

where \(\Gamma( \cdot )\) is the gamma function. To further enhance
numerical stability and handle the zero-inflation probability \(\pi\),
we utilized the softplus function,
\(softplus(x) = log\left( 1 + e^{x} \right)\), and parameterized the
zero-inflation component using logits (\(\rho\)) such that
\(\pi = \text{sigmoid}(\rho) = \frac{1}{1 + e^{- \rho}}\). In our
implementation, we directly predicted the zero-inflation logits
(\(\rho\)), denoted as \texttt{zero\_logits} in our model outputs. The
total loss for a batch of observations was computed as the means of the
individual ZINB losses across all data points in the batch.

Prior to applying the ZINB loss, we performed total count normalization
on the input count data. For each sample, we calculated the sum of all
counts and scaled each count such that the total sum for each sample was
normalized to a target value of \(10^{4}\). This normalization step,
implemented as:
\({\widehat{y}}_{ij} = y_{ij} \times \frac{10^{4}}{\sum_{j}^{}y_{ij}}\),
where \(y_{ij}\) is the original count for feature \(j\) in sample
\(i\), and \({\widehat{y}}_{ij}\) is the normalized count. This step
mitigates the effect of varying sequencing depths across samples,
ensuring fair comparison and model training.

\textbf{Adversarial Network}

To eliminate assay-specific and batch-specific biases in feature
representations, we integrate an adversarial training framework. This
framework employs a~MLP discriminator~and a~gradient reversal layer
(GRL)~\textsuperscript{46} to adversarially optimize the feature
generator. The discriminator is trained to classify assay/batch labels
from the input features, while the generator learns to confound these
predictions via GRL-based gradient inversion. A dynamic loss scaling
strategy further refines the adversarial objective, prioritizing bias
removal as training progresses. This dual adversarial mechanism ensures
robust, assay/batch-invariant representations for downstream
tasks\textsuperscript{47}.

\emph{Adversarial Discriminator Architecture.} We employed a MLP as the
adversarial discriminator, denoted as \(D\). This discriminator network
is designed to classify the domain or assay of the input feature
representations. The discriminator \(D\) consists of \(n_{layers}\)
layers. The discriminator \(D(h)\) is computed through a series of
transformations. Let \(h\) be the input feature representation,
\(h_{1} = W_{1}h + b_{1}\).

For \(i = 2,3,...,n_{layers}\):
\[h_{i}' = \text{LayerNorm}\left( h_{i - 1} \right)\]
\[h_{i}'' = W_{i}h_{i}' + b_{i}\]
\[h_{i}''' = \text{Activation}\left( h_{i}'' \right)\]
\[h_{i} = \text{Dropout}\left( h_{i}''' \right)\]

Finally, the output layer is:

\[D(h) = W_{out}h_{n_{layers}} + b_{out}\]

Where \(W_{i}\) and \(b_{i}\) are the weights and biases of the \(i\)-th
linear layer, respectively. \(\text{LayerNorm}\) represents Layer
Normalization, \(\text{Activation}\) is a non-linear activation function
(LeakyReLU), and \(\text{Dropout}\) is applied with a probability of
0.3. The dimensions of the weight matrices are configured to achieve a
hidden dimension of \(d_{model} \times mlp\_ ratio\). The final linear
layer projects to \(n_{cls}\) output classes, where \(n_{cls}\)
represents the number of domains or assays to be discriminated against.

\emph{Gradient Reversal Layer.} To facilitate adversarial training, a
GRL was inserted before the input to the discriminator. The GRL acts as
an identity function during the forward pass but reverses the gradient
by multiplying it by \(- \lambda\) during backpropagation. Formally, for
an input \(x\), the GRL operation \(GRL(x)\) and its gradient behavior
are defined as:

\begin{quote}
Forward pass: \(GRL(x) = x\)

Backward pass:
\(\frac{\partial L}{\partial x} = - \lambda\frac{\partial L}{\partial y}\)
\end{quote}

where \(y = GRL(x)\) and \(\frac{\partial L}{\partial y}\) is the
gradient from subsequent layers. \(\lambda\) is a hyperparameter
controlling the strength of gradient reversal.

\emph{Adversarial Loss Functions.} Cross-entropy loss is used as the
objective function for both domain and assay adversarial tasks. For
domain adversarial training, the objective is to minimize the
discriminator's ability to correctly identify the domain, thus
encouraging domain-invariant feature learning in the main network. The
adversarial domain loss \(L_{adv\_ domain}\) is defined as:
\(L_{adv\_ domain} = L_{CE}\left( D_{domain}\left( GRL(h) \right),y_{domain} \right)\),
where \(L_{CE}\) is the cross-entropy loss function, \(D_{domain}\) is
the domain discriminator, \(h\) is the feature representation, and
\(y_{domain}\) represents the domain labels. This loss is scaled by a
factor \(\alpha_{adv\_ domain}\) to adjust its contribution to the total
loss.

For assay adversarial training, the goal is to remove assay-specific
biases from the feature representations. The adversarial assay loss
\(L_{adv\_ assay}\) is defined similarly:
\(L_{adv\_ assay} = L_{CE}\left( D_{assay}\left( GRL(h) \right),y_{assay} \right)\),
where \(D_{assay}\) is the assay discriminator and \(y_{assay}\)
represents the assay labels. This loss is scaled by
\(\alpha_{adv\_ assay}\) and a dynamic scaling factor \(s_{epoch}\) that
varies with the training epoch. The dynamic scaling factor \(s_{epoch}\)
is defined as:

\[s_{epoch} = \left\{ \begin{matrix}
0.0001 \times epoch & if\ step < 10000 \\
1 & otherwise
\end{matrix} \right.\ \]

This epoch-dependent scaling progressively increases the influence of
the assay adversarial loss during training.

The total loss function \(L_{total}\) is a weighted sum of the primary
task loss \(L_{main}\), the adversarial domain loss, and the adversarial
assay loss:

\[L_{total} = L_{main} + \alpha_{adv\_ domain}{s_{epoch}L}_{adv\_ domain} + \alpha_{adv\_ assay}s_{epoch}L_{adv\_ assay}\]

By minimizing \(L_{total}\), the model is trained to learn feature
representations that are effective for the primary task while
simultaneously being invariant to domain and assay variations, enhancing
the model's generalization capability and robustness.

\textbf{Class Imbalance Adjustment}

To counteract potential bias arising from class imbalance in the
training data, we implemented a class-balanced weighting scheme based on
the effective number of samples\textsuperscript{48}. Let \(n_{c}\)
denote the number of training samples for class \(c\). The weight
\(w_{c}\) assigned to each class was calculated as:
\(w_{c} = \frac{1 - \beta}{1 - \beta^{n_{c}}}\), where the
hyperparameter \(\beta\) was set to 0.9, following ref. 1. This approach
assigns higher weights to classes with fewer samples.

The computed weights were subsequently normalized to ensure their mean
is unity: \(w_{c,norm} = \frac{w_{c}}{\sum_{i = 1}^{C}w_{i}} \times C\),
where \(C\) is the total number of classes. These normalized weights
\(w_{c,norm}\) were then used to scale the contribution of each class to
the loss function during model training.

\textbf{Classification Loss.}

To address class imbalance and prioritize learning from challenging
examples, we utilized the Focal Loss function¹ as the training
objective. Focal Loss\textsuperscript{49} adapts the standard
cross-entropy loss by incorporating a modulating factor based on the
predicted probability of the true class.

Given the raw output logits \(\mathbf{z}\) from the model for a sample,
we first compute the vector of probabilities
\(p = softmax\left( \mathbf{z} \right)\). Let \(p_{t}\) be the predicted
probability for the ground-truth class \(t\). The Focal Loss (FL) is
defined as:

\[FL\left( p_{t} \right) = - \left( 1 - p_{t} \right)^{\gamma}\log\left( p_{t} \right)\]

where \(\gamma\) is the focusing parameter, set to \(\gamma = 2.0\) in
our study. This formulation down-weights the loss contribution from
easily classified samples (where \(p_{t}\) is high), thereby increasing
the relative importance of misclassified or low-confidence samples.

Computationally, we applied the log-softmax function to the input logits
to obtain log-probabilities. The log-probability corresponding to the
target class, \(\log\left( p_{t} \right)\), was then selected based on
the target class index. The probability \(p_{t}\) was recovered via
exponentiation.

To further account for class frequencies, we incorporated an
alpha-weighting factor,
\(\alpha_{t}\):\(FL\left( p_{t} \right) = - \alpha_{t}\left( 1 - p_{t} \right)^{\gamma}\log\left( p_{t} \right)\).
The \(\alpha_{t}\) values used were the normalized class weights derived
from the effective number of samples strategy (detailed previously). For
each sample in a batch, the appropriate \(\alpha_{t}\) weight
corresponding to its ground-truth class was applied. The final loss
value for a training batch was computed as the arithmetic mean of the
individual focal loss values across all samples within that batch.

\textbf{Gene Expression Loss Function.} The core of the reconstruction
loss is the \texttt{GX\_loss} function, denoted as \(L_{GX}\). This
function, quantifies the difference between the predicted gene
expression distribution and the target gene expression. Let
\(\mathcal{L}_{GX}\left( \widehat{\mathbf{x}},\mathbf{x}_{pk} \right)\)
represent the gene expression loss between the model's output
distribution parameters, summarized as \(\widehat{\mathbf{x}}\), and the
target gene expression profile \(\mathbf{x}_{pk}\). The specific form of
\(\mathcal{L}_{GX}\) is determined by the configuration and may
represent various statistical distances or likelihoods depending on the
chosen gene expression model (e.g., Zero-Inflated Negative Binomial).

\textbf{Loss Calculation.} The loss is computed by differentiating
between masked and unmasked genes based on a mask \(M_{all\_ flat}\).
Let \(M_{all\_ flat}\) be a binary mask indicating which genes are
masked. The gene expression loss \(L_{\exp}\) is then calculated as a
weighted sum of the loss for masked genes and unmasked genes:

\[L_{\exp} = \alpha_{expr}\left( \mathcal{L}_{GX}\left( \widehat{\mathbf{x}},\mathbf{x}_{pk} \right)\left|_{M_{all\_ flat}} + \mathcal{L}_{GX}\left( \widehat{\mathbf{x}},\mathbf{x}_{pk} \right) \right|_{\neg M_{all\_ flat}} \right)\]

where \(\alpha_{expr}\) is a scaling factor controlling the contribution
of the expression reconstruction loss to the total loss.
\(\mathcal{L}_{GX}\left( \widehat{\mathbf{x}},\mathbf{x}_{pk} \right)|_{M_{all\_ flat}}\)
denotes the mean of the gene expression loss evaluated only over the
masked genes (where \(M_{all\_ flat}\) is true), and
\(\mathcal{L}_{GX}\left( \widehat{\mathbf{x}},\mathbf{x}_{pk} \right)|_{\neg M_{all\_ flat}}\)
is the mean loss over the unmasked genes (where \(M_{all\_ flat}\) is
false).

In addition to the gene expression loss \(L_{GX}\), we also monitored
the Mean Squared Error (MSE) between the predicted mean expression and
the target expression for both masked and unmasked genes as diagnostic
metrics, although MSE itself is not directly used as the optimization
objective:\(MSE_{masked} = \text{Mean}\left( \left( {\widehat{\mathbf{\mu}}}_{masked} - \mathbf{x}_{pk,masked} \right)^{2} \right)\)
and
\(MSE_{unmasked} = \text{Mean}\left( \left( {\widehat{\mathbf{\mu}}}_{unmasked} - \mathbf{x}_{pk,unmasked} \right)^{2} \right)\),
where \(\widehat{\mathbf{\mu}}\) represents the predicted mean
expression from the model output, and subscripts \(masked\) and
\(unmasked\) indicate the regions defined by \(M_{all\_ flat}\).

\textbf{Accuracy Metric.} To evaluate the performance of the
classification task, we computed the classification accuracy. Accuracy
is defined as the proportion of correctly classified samples out of the
total number of samples.

\textbf{Handling of Missing Labels.} During training, some samples may
have missing or invalid cell type labels, indicated by a label value of
-1 in our implementation. To ensure that these samples do not contribute
to the classification loss, we filtered out samples with labels equal to
-1 before computing the cross-entropy loss and accuracy. Specifically,
we only considered samples where \(cell\_ labels \neq - 1\) for loss
calculation and accuracy evaluation.

\textbf{Loss Scaling.} The classification loss was scaled by a factor
\(\alpha_{cls}\) to adjust its contribution to the total loss, allowing
for fine-tuning the balance between different loss terms if combined
with other objectives (e.g., adversarial losses). In our experiments,
the classification loss scale was set to 1.0 by default unless otherwise
specified.

The minimization of \(L_{cls}\) drives the model to learn feature
representations that are discriminative for different cell types,
enabling accurate classification of cells based on their learned
representations.

\subsubsection{\texorpdfstring{\textbf{Training
Procedure}}{Training Procedure}}\label{training-procedure}

We employed a Fully Sharded Data Parallel training strategy across 4
H100 GPUs to accelerate the training process. The model was trained
end-to-end, minimizing a combined loss function that incorporates both
gene expression reconstruction and cell type classification objectives,
and optionally adversarial domain and assay adaptation losses.

\emph{Optimization Algorithm.} We used the Adam or AdamW optimizer to
update the model parameters. The optimizer was configured with an
initial learning rate (\(\eta\)), and optionally a weight decay
(\(\lambda\)) for regularization.

\[\theta_{t + 1} = \text{Optimizer}\left( \theta_{t},\nabla L_{total}\left( \theta_{t} \right),\eta,\lambda \right)\]

where \(\theta_{t}\) represents the model parameters at training step
\(t\), and \(\nabla L_{total}\left( \theta_{t} \right)\) is the gradient
of the total loss with respect to the parameters.

\emph{Learning Rate Scheduling.} A learning rate scheduler was employed
to adjust the learning rate during training. We utilized either a
ReduceLROnPlateau scheduler, which reduces the learning rate when
validation loss plateaus, or a CosineAnnealingLR scheduler, which
follows a cosine annealing schedule. For ReduceLROnPlateau, the learning
rate is updated based on validation loss \(L_{val}\):

\[\eta_{t + 1} = \left\{ \begin{matrix}
\eta_{t} \times \text{factor} & \text{if }L_{val}\text{ plateaus} \\
\eta_{t} & \text{otherwise}
\end{matrix} \right.\ \]

For CosineAnnealingLR, the learning rate follows a cosine function over
training steps.

\emph{Learning Rate Warmup.} To stabilize initial training, a linear
learning rate warmup strategy was implemented for the first
\(N_{warmup}\) steps. During warmup, the learning rate \(\eta_{t}\) at
step \(t\) is:

\[\eta_{t} = \eta_{initial} + \left( \eta - \eta_{initial} \right) \times \frac{t}{N_{warmup}}\]

where \(\eta_{initial}\) is a small initial learning rate (effectively 0
in our setup, starting from a very small value) and \(\eta\) is the
target learning rate.

\emph{Gradient Clipping.} To prevent exploding gradients, we applied
gradient clipping by norm. The gradients were clipped such that their L2
norm does not exceed a predefined threshold (e.g., 1.0).

\[\mathbf{g}' = \left\{ \begin{matrix}
\frac{\text{clip\_norm}}{\parallel \mathbf{g} \parallel_{2}}\mathbf{g} & \text{if } \parallel \mathbf{g} \parallel_{2} > \text{clip\_norm} \\
\mathbf{g} & \text{otherwise}
\end{matrix} \right.\ \]

where \(\mathbf{g}\) is the gradient vector, \(\mathbf{g}'\) is the
clipped gradient vector, and \(\text{clip\_norm}\) is the clipping
threshold.

\emph{Mixed Precision Training.} To accelerate training and reduce
memory consumption, we used Automatic Mixed Precision (AMP) training via
\texttt{torch.cuda.amp.GradScaler} and \texttt{torch.cuda.amp.autocast}.
This technique performs computations in half-precision (float16) where
possible, while maintaining gradients and parameter updates in full
precision (float32) for stability.

\emph{Model Initialization.} Model parameters were initialized using
Xavier uniform initialization for linear layers, and biases were
initialized to zero.

\textbf{Data Availability}

The previously published data generated for this study are available in
GSE107585, GSE182256, GSE183842, GSE173343, GSE211785, GSE209821,
GSE183839, and GSE291551. Raw data, processed data, and metadata from
the scRNA-seq and CosMx spatial transcriptomics experiments have been
deposited in the Gene Expression Omnibus (GEO) under accession code ***,
with reviewer token ***.

\textbf{Acknowledgments}

The authors would like to acknowledge Dylan Jay, David Render, Sean
LePeruta, Jovana Lekic, Pio Passariello, and Hannah Hollosi for their
valuable contributions and support to this study.

\textbf{Authors\textquotesingle{} Contribution}

Concept and design: CL, MS, NZ and KS; Manuscript drafting: CL, KS;
Statistical analysis: CL and EZ; Data collection and interpretation: CL,
EZ, BD, JL, EH and SP; Validation: EZ, YS, VR and MS; Manuscript
revision: CL, NZ and KS. All authors critically reviewed and agreed to
the submission of the final manuscript.

\textbf{Funding}

This work was supported by the National Institutes of Health grant
National Institute of Diabetes and Digestive and Kidney Diseases (NIDDK)
R01 DK076077, R01 DK087635, and R01 DK105821 (to KS).
\end{doublespace}
\textbf{Reference}

1. Ramesh A, Dhariwal P, Nichol A, Chu C and Chen M. Hierarchical
Text-Conditional Image Generation with CLIP Latents: arXiv:2204.06125
(2022, accessed April 01, 2022).

2. Brown TB, Mann B, Ryder N, et al. Language Models are Few-Shot
Learners: arXiv:2005.14165 (2020, accessed May 01, 2020).

3. Cui H, Wang C, Maan H, et al. scGPT: toward building a foundation
model for single-cell multi-omics using generative AI. Nat Methods.
2024; 21: 1470-80.

4. Theodoris CV, Xiao L, Chopra A, et al. Transfer learning enables
predictions in network biology. Nature. 2023; 618.

5. Hao MS, Gong J, Zeng X, et al. Large-scale foundation model on
single-cell transcriptomics. Nature Methods. 2024; 21.

6. Moor M, Banerjee O, Abad ZSH, et al. Foundation models for generalist
medical artificial intelligence. Nature. 2023; 616: 259-65.

7. Richter T, Bahrami M, Xia YF, Fischer DS and Theis FJ. Delineating
the effective use of self-supervised learning in single-cell genomics.
Nat Mach Intell. 2025; 7.

8. Lake BB, Menon R, Winfree S, et al. An atlas of healthy and injured
cell states and niches in the human kidney. Nature. 2023; 619: 585-94.

9. Gerhardt LMS, Koppitch K, van Gestel J, et al. Lineage Tracing and
Single-Nucleus Multiomics Reveal Novel Features of Adaptive and
Maladaptive Repair after Acute Kidney Injury. J Am Soc Nephrol. 2023;
34: 554-71.

10. Muto Y, Wilson PC, Ledru N, et al. Single cell transcriptional and
chromatin accessibility profiling redefine cellular heterogeneity in the
adult human kidney. Nat Commun. 2021; 12: 2190.

11. Kirita Y, Wu H, Uchimura K, Wilson PC and Humphreys BD. Cell
profiling of mouse acute kidney injury reveals conserved cellular
responses to injury. Proc Natl Acad Sci U S A. 2020; 117: 15874-83.

12. Romagnani P, Agarwal R, Chan JCN, et al. Chronic kidney disease. Nat
Rev Dis Primers. 2025; 11: 8.

13. Desanti De Oliveira B, Xu K, Shen TH, et al. Molecular nephrology:
types of acute tubular injury. Nat Rev Nephrol. 2019; 15: 599-612.

14. Stewart BJ, Ferdinand JR, Young MD, et al. Spatiotemporal immune
zonation of the human kidney. Science. 2019; 365: 1461-6.

15. Park J, Shrestha R, Qiu C, et al. Single-cell transcriptomics of the
mouse kidney reveals potential cellular targets of kidney disease.
Science. 2018; 360: 758-63.

16. Hao Y, Stuart T, Kowalski MH, et al. Dictionary learning for
integrative, multimodal and scalable single-cell analysis. Nat
Biotechnol. 2024; 42: 293-304.

17. Wolf FA, Angerer P and Theis FJ. SCANPY: large-scale single-cell
gene expression data analysis. Genome Biol. 2018; 19: 15.

18. Hansen J, Sealfon R, Menon R, et al. A reference tissue atlas for
the human kidney. Sci Adv. 2022; 8: eabn4965.

19. Luecken MD, Buttner M, Chaichoompu K, et al. Benchmarking
atlas-level data integration in single-cell genomics. Nat Methods. 2022;
19: 41-50.

20. Stuart T, Butler A, Hoffman P, et al. Comprehensive Integration of
Single-Cell Data. Cell. 2019; 177: 1888-902 e21.

21. Theodoris CV, Xiao L, Chopra A, et al. Transfer learning enables
predictions in network biology. Nature. 2023; 618: 616-24.

22. Hao M, Gong J, Zeng X, et al. Large-scale foundation model on
single-cell transcriptomics. Nat Methods. 2024; 21: 1481-91.

23. Roohani Y, Huang K and Leskovec J. Predicting transcriptional
outcomes of novel multigene perturbations with GEARS. Nat Biotechnol.
2024; 42: 927-35.

24. Jager KJ, Kovesdy C, Langham R, Rosenberg M, Jha V and Zoccali C. A
single number for advocacy and communication-worldwide more than 850
million individuals have kidney diseases. Nephrol Dial Transplant. 2019;
34: 1803-5.

25. Abedini A, Levinsohn J, Klotzer KA, et al. Single-cell multi-omic
and spatial profiling of human kidneys implicates the fibrotic
microenvironment in kidney disease progression. Nat Genet. 2024; 56:
1712-24.

26. Balzer MS, Doke T, Yang YW, et al. Single-cell analysis highlights
differences in druggable pathways underlying adaptive or fibrotic kidney
regeneration. Nat Commun. 2022; 13: 4018.

27. Edgar R, Domrachev M and Lash AE. Gene Expression Omnibus: NCBI gene
expression and hybridization array data repository. Nucleic Acids Res.
2002; 30: 207-10.

28. Regev A, Teichmann SA, Lander ES, et al. The Human Cell Atlas.
Elife. 2017; 6.

29. Program CS-CB, Abdulla S, Aevermann B, et al. CZ CELL×GENE Discover:
A single-cell data platform for scalable exploration, analysis and
modeling of aggregated data. bioRxiv. 2023: 2023.10.30.563174.

30. Wu H, Dixon EE, Xuanyuan Q, et al. High resolution spatial profiling
of kidney injury and repair using RNA hybridization-based in situ
sequencing. Nat Commun. 2024; 15: 1396.

31. Harrison PW, Amode MR, Austine-Orimoloye O, et al. Ensembl 2024.
Nucleic Acids Res. 2024; 52: D891-D9.

32. Fleming SJ, Chaffin MD, Arduini A, et al. Unsupervised removal of
systematic background noise from droplet-based single-cell experiments
using CellBender. Nat Methods. 2023; 20: 1323-35.

33. Zhang B and Sennrich R. Root Mean Square Layer Normalization:
arXiv:1910.07467 (2019, accessed October 01, 2019).

34. Schwarz A, Sklyar I and Wiesler S. Improving RNN-T ASR Accuracy
Using Context Audio: arXiv:2011.10538 (2020, accessed November 01,
2020).

35. Ramachandran P, Zoph B and Le QV. Searching for Activation
Functions: arXiv:1710.05941 (2017, accessed October 01, 2017).

36. Vinogradova K, Dibrov A and Myers G. Towards Interpretable Semantic
Segmentation via Gradient-weighted Class Activation Mapping:
arXiv:2002.11434 (2020, accessed February 01, 2020).

37. Shazeer N, Mirhoseini A, Maziarz K, et al. Outrageously Large Neural
Networks: The Sparsely-Gated Mixture-of-Experts Layer: arXiv:1701.06538
(2017, accessed January 01, 2017).

38. Fedus W, Zoph B and Shazeer N. Switch Transformers: Scaling to
Trillion Parameter Models with Simple and Efficient Sparsity:
arXiv:2101.03961 (2021, accessed January 01, 2021).

39. Lepikhin D, Lee H, Xu Y, et al. GShard: Scaling Giant Models with
Conditional Computation and Automatic Sharding: arXiv:2006.16668 (2020,
accessed June 01, 2020).

40. Dai D, Deng C, Zhao C, et al. DeepSeekMoE: Towards Ultimate Expert
Specialization in Mixture-of-Experts Language Models: arXiv:2401.06066
(2024, accessed January 01, 2024).

41. Khosla P, Teterwak P, Wang C, et al. Supervised Contrastive
Learning: arXiv:2004.11362 (2020, accessed April 01, 2020).

42. Chen T, Kornblith S, Norouzi M and Hinton G. A Simple Framework for
Contrastive Learning of Visual Representations: arXiv:2002.05709 (2020,
accessed February 01, 2020).

43. van den Oord A, Li Y and Vinyals O. Representation Learning with
Contrastive Predictive Coding: arXiv:1807.03748 (2018, accessed July 01,
2018).

44. Lopez R, Regier J, Cole MB, Jordan MI and Yosef N. Deep generative
modeling for single-cell transcriptomics. Nat Methods. 2018; 15: 1053-8.

45. Kalfon J, Samaran J, Peyre G and Cantini L. scPRINT: pre-training on
50 million cells allows robust gene network predictions. Nat Commun.
2025; 16: 3607.

46. Goodfellow IJ, Pouget-Abadie J, Mirza M, et al. Generative
Adversarial Networks: arXiv:1406.2661 (2014, accessed June 01, 2014).

47. Yang KD, Belyaeva A, Venkatachalapathy S, et al. Multi-domain
translation between single-cell imaging and sequencing data using
autoencoders. Nat Commun. 2021; 12: 31.

48. Cui Y, Jia M, Lin T-Y, Song Y and Belongie S. Class-Balanced Loss
Based on Effective Number of Samples: arXiv:1901.05555 (2019, accessed
January 01, 2019).

49. Lin TY, Goyal P, Girshick R, He K and Dollar P. Focal Loss for Dense
Object Detection. IEEE Trans Pattern Anal Mach Intell. 2020; 42: 318-27.
\newpage
\includegraphics[width=\linewidth]{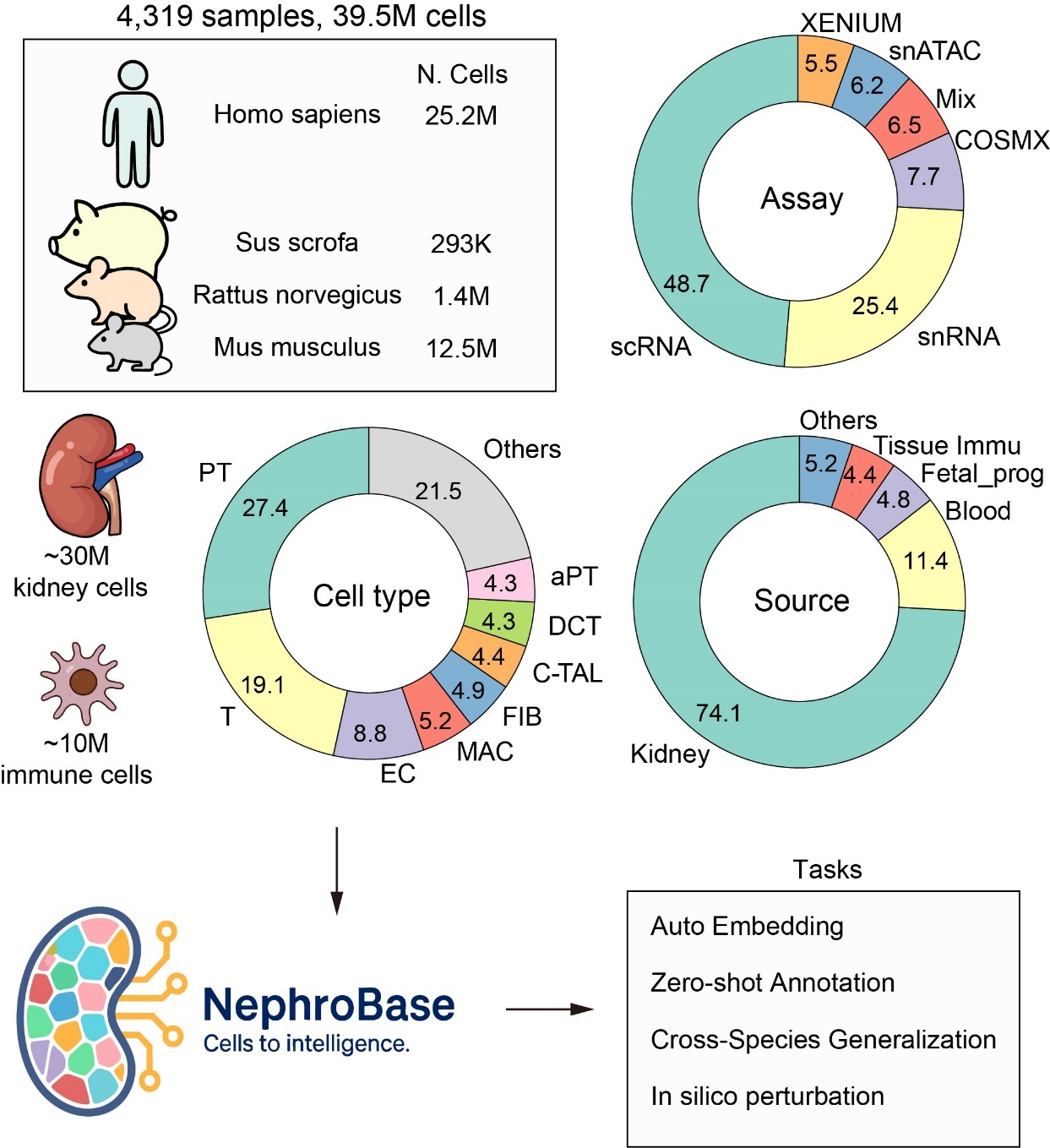}

\textbf{Figure 1. Composition of the Nephrobase Cell+ training dataset.}

The Nephrobase Cell+ atlas integrates 4,319 samples encompassing
\textasciitilde39.5 million single-cell and single-nucleus profiles. Top
left: Distribution of cells across four mammalian species: Homo sapiens
(25.2M), Mus musculus (12.5M), Rattus norvegicus (1.4M), and Sus scrofa
(0.293M). Bottom left: Approximate tissue origin of the dataset,
including \textasciitilde30M kidney-derived cells and \textasciitilde10M
immune cells. The cell-type composition shows strong representation of
proximal tubule (PT, 27.4\%), T cells (19.1\%), endothelial cells (EC,
8.8\%), macrophages (MAC, 5.2\%), fibroblasts (FIB, 4.9\%), cortical
thick ascending limb (C-TAL, 4.3\%), distal convoluted tubule (DCT,
4.3\%), and atrophic proximal tubule (aPT, 4.3\%), with the remainder
categorized as ``others'' (21.5\%). Top right: Assay composition,
highlighting contributions from scRNA-seq (48.7\%), snRNA-seq (25.4\%),
COSMx (7.7\%), Xenium (5.5\%), snATAC-seq (6.2\%), and mixed modalities
(6.5\%). Bottom right: Sample source composition, showing that most
profiles are from kidney tissue (74.1\%), with additional contributions
from blood (11.4\%), fetal/progenitor samples (4.8\%), tissue-enriched
immune fractions (4.4\%), and other sources (5.2\%). Together, these
distributions illustrate the multimodal, multispecies, and multicontext
diversity of the Nephrobase Cell+ training dataset.
\newpage
\includegraphics[width=\linewidth]{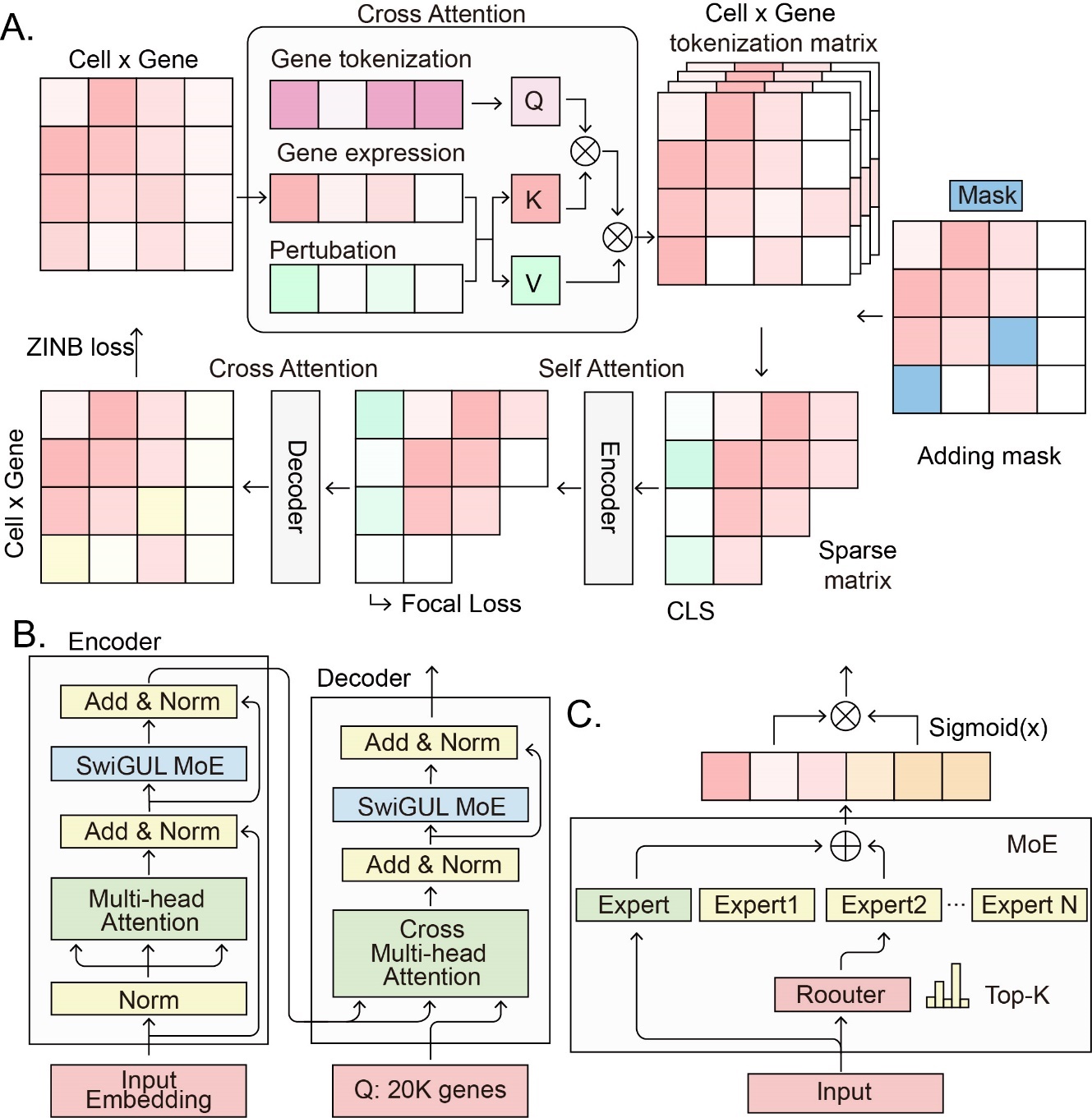}

\textbf{Figure 2. Nephrobase Cell+ model architecture and training
strategy.} (A) Overview of the encoder-decoder framework. The model
ingests a cell-by-gene matrix, with each gene tokenized using its
identity, normalized expression value, and optional perturbation
metadata. Tokens are embedded via cross-attention to generate a cell ×
gene tokenization matrix, with masking applied to subsets of inputs. The
encoder applies self-attention, while the decoder uses cross-attention
to reconstruct expression profiles. Outputs are optimized using a
Zero-Inflated Negative Binomial (ZINB) loss for count reconstruction and
a focal loss for supervised cell-type classification. (B) Detailed
transformer block design. Both encoder and decoder stacks include
normalization layers, multi-head attention modules, and SwiGUL
Mixture-of-Experts (MoE) layers, with cross multi-head attention
connecting the decoder to the encoder. Input embeddings represent up to
20,000 genes per cell. (C) Structure of the Mixture-of-Experts (MoE)
module. Each input is routed to a subset of specialized experts using
top-k gating, with outputs combined through weighted summation. A shared
expert and sigmoid activation further stabilize and generalize
representation learning. Together, these components allow Nephrobase
Cell+ to learn robust, assay-invariant embeddings of kidney cell states
from large-scale multi-modal data.
\newpage
\includegraphics[width=\linewidth]{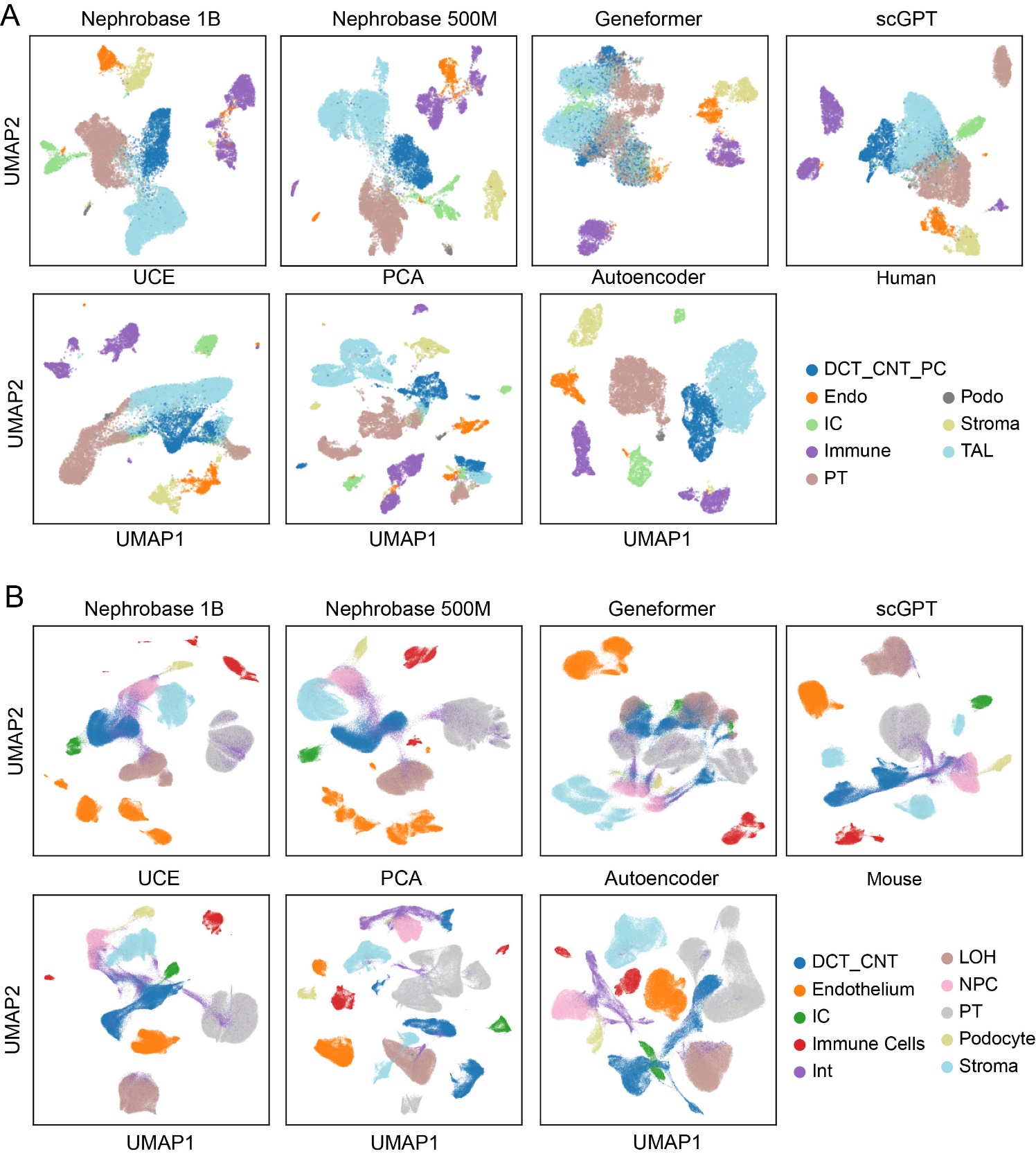}

\textbf{Figure 3. Zero-shot benchmarking of embedding of single-nucleus
transcriptomic data analysis methods using for human and mouse
datasets.}(A) UMAP projections of human kidney data from various
single-cell RNA-seq analysis methods: Nephrobase Cell+ 1B, Nephrobase
Cell+ 500M, Genformer, and scGPT. Each method is shown with clustering
using Uniformed Cluster Embedding (UCE), Principal Component Analysis
(PCA), and Autoencoder dimensionality reduction techniques. Proximal
Tubule (PT) (cyan), Stroma (yellow), and TAL (light blue). (B) UMAP
projections of mouse kidney data following the same methods and
dimensionality reduction techniques as in (A). DCT\_CNT\_PC (Distal
Convoluted Tubule and Connecting Tubule Principal Cells), Endo
(Endothelium), IC (Intercalated Cells), Immune (Immune Cells), Podo
(Podocytes), PT (Proximal Tubule), Stroma (Stromal Cells), TAL (Thick
Ascending Limb), LOH (Loop of Henle), NPC (Nephron Progenitor Cells),
Int (Interstitial Cel
\begin{landscape}
\newpage
\includegraphics[width=\linewidth]{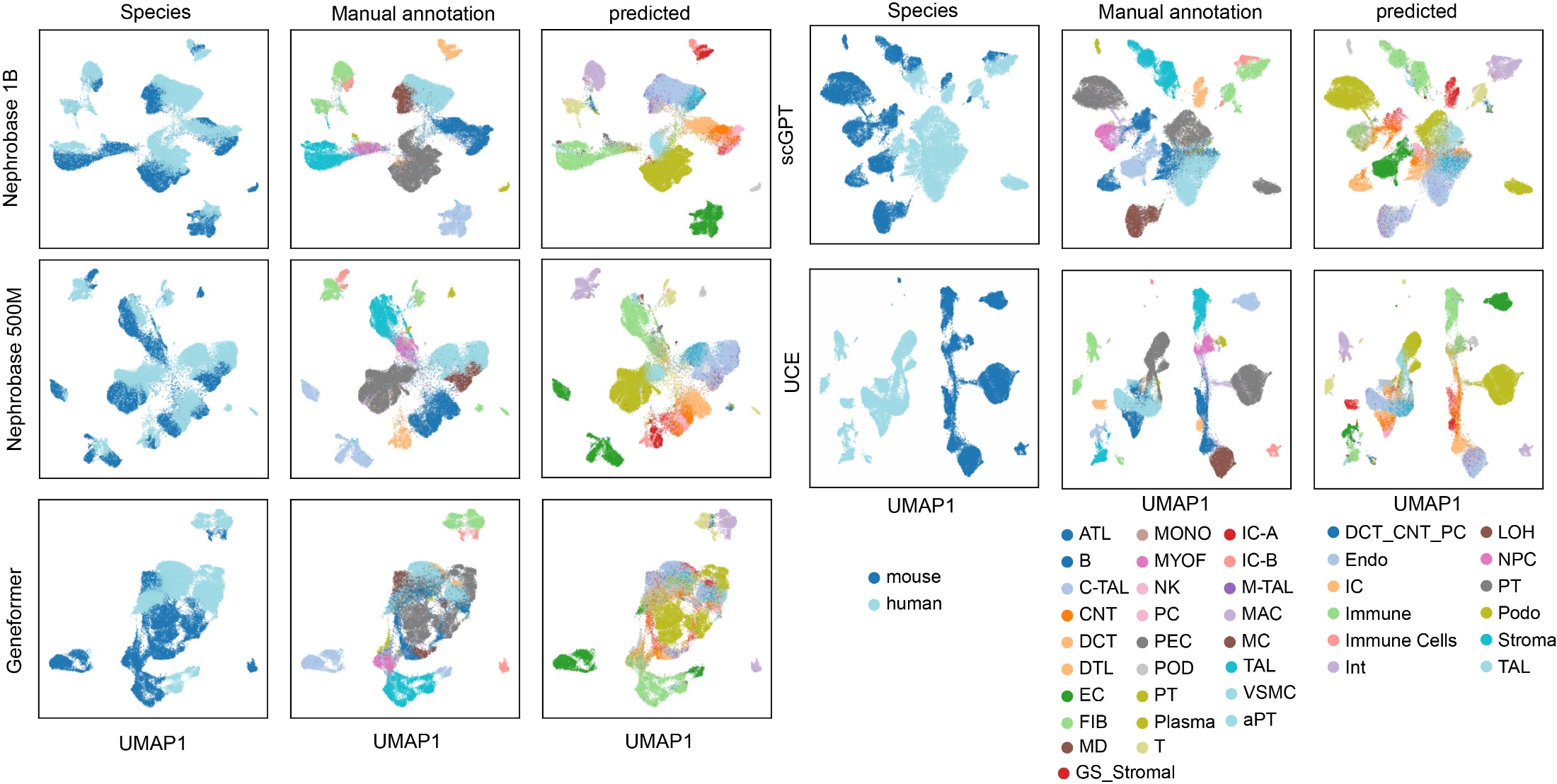}

\textbf{Figure 4. Zero-shot cross-species benchmarking of foundation
models for kidney single-\/-nucleus transcriptomics.}

UMAP visualizations of human and mouse kidney single-cell transcriptomic
data comparing species labels, manual expert annotations, and
model-predicted cell types across multiple foundation models for
Nephrobase Cell+ models (1B and 500M parameters),Geneformer, scGPT and
UCE. Each row corresponds to a distinct model, with columns showing (i)
species distribution (human, light blue; mouse, dark blue), (ii) manual
cell type annotations, and (iii) zero-shot model predictions. Major
kidney epithelial, stromal, endothelial, and immune cell types are
highlighted, including proximal tubule (PT), thick ascending limb (TAL),
distal convoluted tubule/connecting tubule (DCT/CNT), intercalated cells
(IC), podocytes (PODO), stromal cells, endothelial cells (Endo), immune
cells, and nephron progenitors (NPC). Predictions from Nephrobase Cell+
and scGPT more closely recapitulate expert manual annotations compared
to Geneformer and UCE, demonstrating improved cross-species
generalizability and fine-grained nephron cell type resolution.

\end{landscape}

\includegraphics[width=\linewidth]{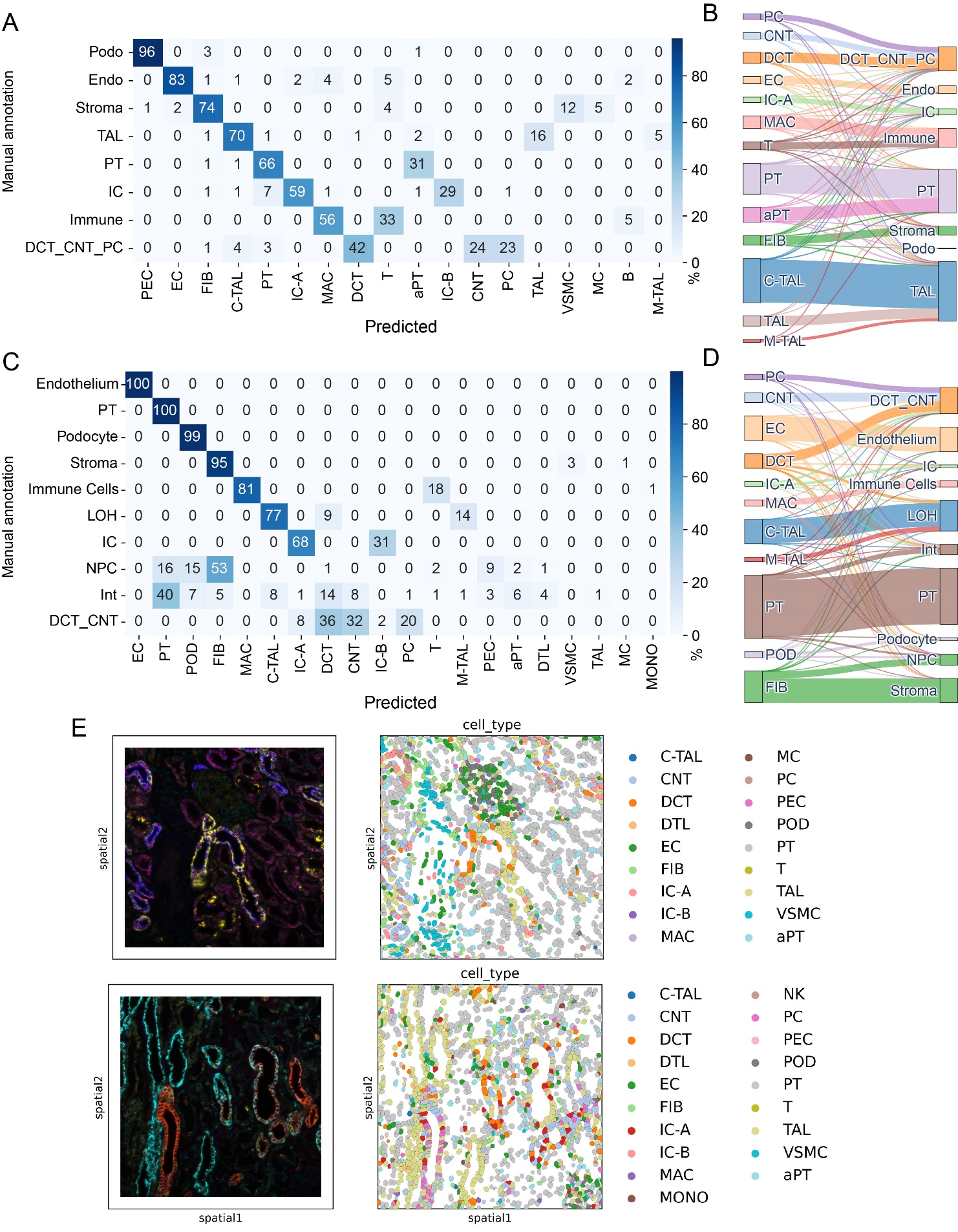}

\textbf{Figure 5. Zero-shot benchmarking Nephrobase Cell+ cell type
annotation against manual curation.}

Confusion matrices showing agreement between manual annotations (rows)
and Nephrobase Cell+ predictions (columns) across major kidney cell
types in (A) human and (C) mouse datasets. The percentage of cells
correctly assigned to each category is indicated by the color intensity,
with darker shades reflecting higher concordance. (B, D) Sankey diagrams
illustrating the mapping between manual annotations and Nephrobase Cell+
predictions for the same datasets. (E) an example should the predict
cell type for spatial transcriptome.Major nephron epithelial lineages
(PT, TAL, DCT, CNT, IC, Podocytes) as well as stromal, endothelial, and
immune compartments are shown. The width of each connection is
proportional to the number of cells assigned. Together, these analyses
demonstrate that Nephrobase Cell+ achieves high concordance with expert
manual curation while capturing fine-grained nephron subtypes.
\newpage
\includegraphics[width=\linewidth]{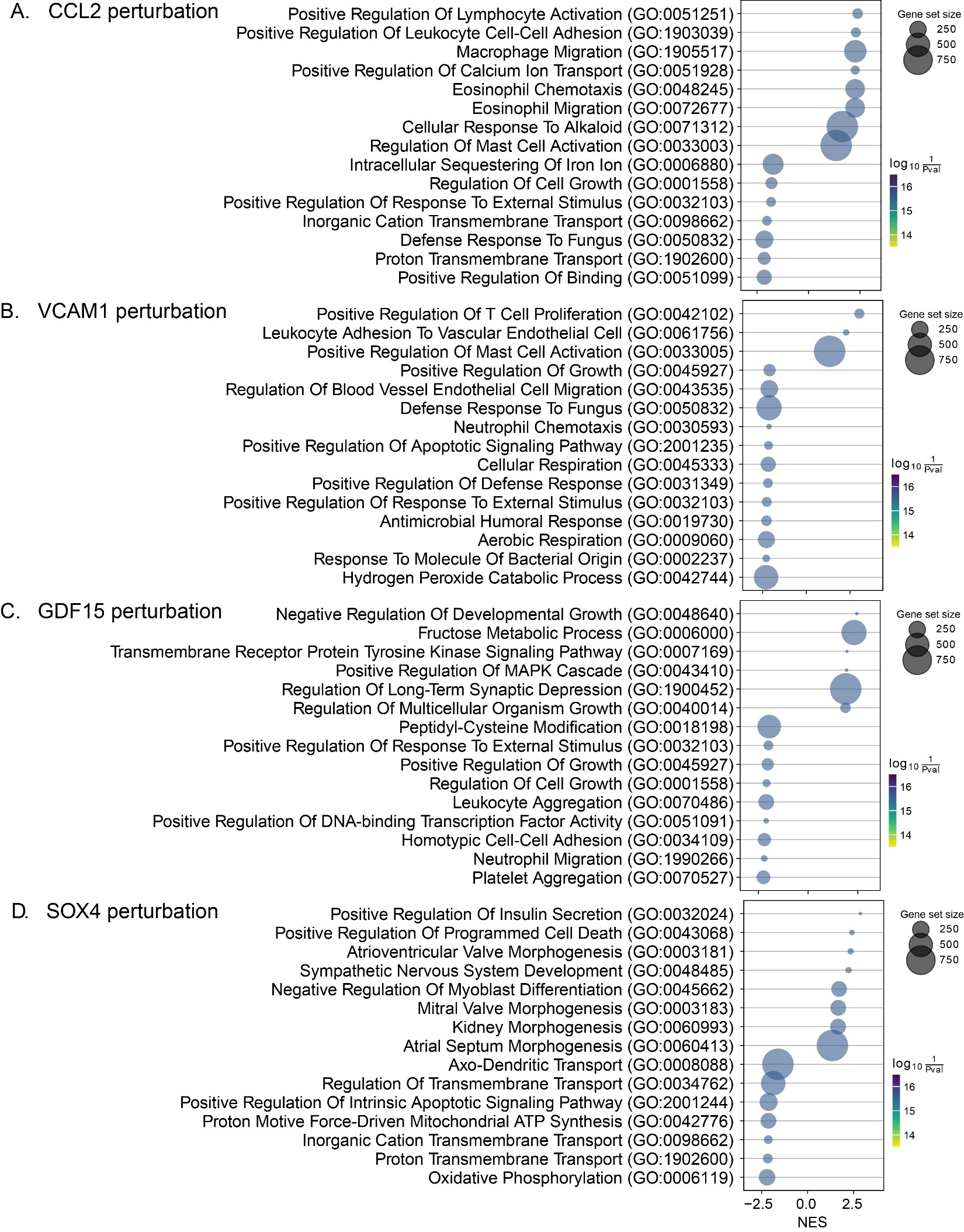}

\textbf{Figure 6. Expression-driven Gene Ontology (GO) enrichment
results for four gene perturbations.} Panels A-D show the top enriched
GO Biological Process terms following perturbation of CCL2 (A), VCAM1
(B), GDF15 (C) and SOX4 (D). Each bubble represents one GO term; the
x-axis shows the normalized enrichment score (NES), bubble size
corresponds to the gene set size (number of genes in the GO term), and
bubble color encodes statistical significance as −log10(p-value) (darker
= more significant). Terms are ordered by significance and effect size
and only the most enriched / interpretable terms are displayed for
clarity. Positive NES values indicate enrichment among up-regulated
genes after perturbation, while negative NES values indicate enrichment
among down-regulated genes. Enrichment was calculated using gene set
enrichment analysis (GSEA) on ranked differential expression results,
and GO terms shown are from the Biological Process ontology.
\newpage
\textbf{Table 1. Model Architecture and Training Hyperparameters of
Nephrobase Cell+ Variants*}

\begin{longtable}[]{@{}
  >{\centering\arraybackslash}p{(\linewidth - 4\tabcolsep) * \real{0.3333}}
  >{\centering\arraybackslash}p{(\linewidth - 4\tabcolsep) * \real{0.3333}}
  >{\centering\arraybackslash}p{(\linewidth - 4\tabcolsep) * \real{0.3333}}@{}}
\toprule\noalign{}
\begin{minipage}[b]{\linewidth}\centering
\end{minipage} & \begin{minipage}[b]{\linewidth}\centering
\textbf{Nephrobase Cell+ 1B}
\end{minipage} & \begin{minipage}[b]{\linewidth}\centering
\textbf{Nephrobase Cell+ 500M}
\end{minipage} \\
\midrule\noalign{}
\endhead
\bottomrule\noalign{}
\endlastfoot
embed\_d & 1024 & 768 \\
batch\_size & 90 & 136 \\
enc\_nheads & 8 & 8 \\
dec\_nheads & 8 & 8 \\
n\_fuse\_attention & 1 & 1 \\
n\_encoder & 6 & 5 \\
n\_decoder & 1 & 1 \\
enc\_mlp\_ratio & 4 & 4 \\
dec\_mlp\_ratio & 4 & 4 \\
num\_classes & 31 & 31 \\
Training strategy & FSDP & FSDP \\
Parameter & \textasciitilde500M & \textasciitilde1B \\
\end{longtable}

*embed\_d, embedding dimension; enc\_nheads, number of encoder attention
heads; dec\_nheads, number of decoder attention heads;
n\_fuse\_attention, number of fusion attention layers; n\_encoder,
number of encoder layers; n\_decoder, number of decoder layers;
enc\_mlp\_ratio, encoder multilayer-perceptron ratio; dec\_mlp\_ratio,
decoder multilayer-perceptron ratio; FSDP, Fully Sharded Data Parallel.
\newpage
\textbf{Table 2. Batch effect removing benchmarking Nephrobase Cell+
against existing foundational and dimensionality reduction models in
human and mouse datasets.*}

\begin{longtable}[]{@{}
  >{\centering\arraybackslash}p{(\linewidth - 18\tabcolsep) * \real{0.0769}}
  >{\centering\arraybackslash}p{(\linewidth - 18\tabcolsep) * \real{0.1725}}
  >{\centering\arraybackslash}p{(\linewidth - 18\tabcolsep) * \real{0.0141}}
  >{\centering\arraybackslash}p{(\linewidth - 18\tabcolsep) * \real{0.1235}}
  >{\centering\arraybackslash}p{(\linewidth - 18\tabcolsep) * \real{0.1235}}
  >{\centering\arraybackslash}p{(\linewidth - 18\tabcolsep) * \real{0.1347}}
  >{\centering\arraybackslash}p{(\linewidth - 18\tabcolsep) * \real{0.1299}}
  >{\centering\arraybackslash}p{(\linewidth - 18\tabcolsep) * \real{0.0714}}
  >{\centering\arraybackslash}p{(\linewidth - 18\tabcolsep) * \real{0.0749}}
  >{\centering\arraybackslash}p{(\linewidth - 18\tabcolsep) * \real{0.0785}}@{}}
\toprule\noalign{}
\begin{minipage}[b]{\linewidth}\raggedright
\end{minipage} & \begin{minipage}[b]{\linewidth}\centering
\textbf{Metrics}
\end{minipage} &
\multicolumn{2}{>{\centering\arraybackslash}p{(\linewidth - 18\tabcolsep) * \real{0.1376} + 2\tabcolsep}}{%
\begin{minipage}[b]{\linewidth}\centering
\textbf{Nephrobase Cell+}

\textbf{1B}
\end{minipage}} & \begin{minipage}[b]{\linewidth}\centering
\textbf{Nephrobase Cell+}

\textbf{500M}
\end{minipage} & \begin{minipage}[b]{\linewidth}\centering
\textbf{Autoencoder}
\end{minipage} & \begin{minipage}[b]{\linewidth}\centering
\textbf{Geneformer}
\end{minipage} & \begin{minipage}[b]{\linewidth}\centering
\textbf{UCE}
\end{minipage} & \begin{minipage}[b]{\linewidth}\centering
\textbf{PCA}
\end{minipage} & \begin{minipage}[b]{\linewidth}\centering
\textbf{scGPT}
\end{minipage} \\
\midrule\noalign{}
\endhead
\bottomrule\noalign{}
\endlastfoot
\multirow{13}{=}{\centering\arraybackslash Human} &
\multicolumn{2}{>{\centering\arraybackslash}p{(\linewidth - 18\tabcolsep) * \real{0.1866} + 2\tabcolsep}}{%
Isolated labels} & 0.76 & 0.77 & 0.75 & 0.58 & 0.66 & \textbf{0.89} &
0.62 \\
&
\multicolumn{2}{>{\centering\arraybackslash}p{(\linewidth - 18\tabcolsep) * \real{0.1866} + 2\tabcolsep}}{%
KMeans NMI} & \textbf{0.78} & 0.76 & 0.72 & 0.37 & 0.63 & 0.54 & 0.48 \\
&
\multicolumn{2}{>{\centering\arraybackslash}p{(\linewidth - 18\tabcolsep) * \real{0.1866} + 2\tabcolsep}}{%
KMeans ARI} & \textbf{0.82} & 0.67 & 0.55 & 0.22 & 0.48 & 0.4 & 0.3 \\
&
\multicolumn{2}{>{\centering\arraybackslash}p{(\linewidth - 18\tabcolsep) * \real{0.1866} + 2\tabcolsep}}{%
Silhouette label} & \textbf{0.68} & 0.67 & 0.62 & 0.52 & 0.6 & 0.58 &
0.55 \\
&
\multicolumn{2}{>{\centering\arraybackslash}p{(\linewidth - 18\tabcolsep) * \real{0.1866} + 2\tabcolsep}}{%
cLISI} & \textbf{1} & \textbf{1} & \textbf{1} & 0.97 & \textbf{1} &
\textbf{1} & \textbf{1} \\
&
\multicolumn{2}{>{\centering\arraybackslash}p{(\linewidth - 18\tabcolsep) * \real{0.1866} + 2\tabcolsep}}{%
BRAS} & 0.74 & \textbf{0.77} & 0.79 & 0.77 & 0.72 & 0.28 & 0.71 \\
&
\multicolumn{2}{>{\centering\arraybackslash}p{(\linewidth - 18\tabcolsep) * \real{0.1866} + 2\tabcolsep}}{%
iLISI} & 0.17 & \textbf{0.18} & 0.13 & 0.12 & 0.06 & 0.01 & 0.09 \\
&
\multicolumn{2}{>{\centering\arraybackslash}p{(\linewidth - 18\tabcolsep) * \real{0.1866} + 2\tabcolsep}}{%
KBET} & 0.25 & \textbf{0.28} & 0.1 & 0.12 & 0.07 & 0.09 & 0.1 \\
&
\multicolumn{2}{>{\centering\arraybackslash}p{(\linewidth - 18\tabcolsep) * \real{0.1866} + 2\tabcolsep}}{%
Graph connectivity} & 0.94 & 0.93 & \textbf{0.98} & 0.8 & 0.84 & 0.71 &
0.88 \\
&
\multicolumn{2}{>{\centering\arraybackslash}p{(\linewidth - 18\tabcolsep) * \real{0.1866} + 2\tabcolsep}}{%
PCR comparison} & 0.67 & 0.73 & \textbf{0.84} & 0.44 & 0.38 & 0.13 &
0.36 \\
&
\multicolumn{2}{>{\centering\arraybackslash}p{(\linewidth - 18\tabcolsep) * \real{0.1866} + 2\tabcolsep}}{%
Batch correction} & 0.55 & \textbf{0.58} & 0.57 & 0.45 & 0.41 & 0.24 &
0.43 \\
&
\multicolumn{2}{>{\centering\arraybackslash}p{(\linewidth - 18\tabcolsep) * \real{0.1866} + 2\tabcolsep}}{%
Bio conservation} & \textbf{0.81} & 0.77 & 0.73 & 0.54 & 0.67 & 0.68 &
0.59 \\
&
\multicolumn{2}{>{\centering\arraybackslash}p{(\linewidth - 18\tabcolsep) * \real{0.1866} + 2\tabcolsep}}{%
Total} & \textbf{0.71} & 0.7 & 0.67 & 0.5 & 0.57 & 0.51 & 0.53 \\
\multirow{13}{=}{\centering\arraybackslash Mouse} &
\multicolumn{2}{>{\centering\arraybackslash}p{(\linewidth - 18\tabcolsep) * \real{0.1866} + 2\tabcolsep}}{%
Isolated labels} & 0.64 & 0.62 & 0.77 & 0.58 & 0.65 & \textbf{0.87} &
0.66 \\
&
\multicolumn{2}{>{\centering\arraybackslash}p{(\linewidth - 18\tabcolsep) * \real{0.1866} + 2\tabcolsep}}{%
KMeans NMI} & \textbf{0.79} & \textbf{0.79} & \textbf{0.79} & 0.6 & 0.75
& 0.69 & 0.79 \\
&
\multicolumn{2}{>{\centering\arraybackslash}p{(\linewidth - 18\tabcolsep) * \real{0.1866} + 2\tabcolsep}}{%
KMeans ARI} & \textbf{0.7} & \textbf{0.68} & 0.65 & 0.41 & 0.6 & 0.53 &
0.65 \\
&
\multicolumn{2}{>{\centering\arraybackslash}p{(\linewidth - 18\tabcolsep) * \real{0.1866} + 2\tabcolsep}}{%
Silhouette label} & \textbf{0.69} & 0.66 & 0.63 & 0.55 & \textbf{0.67} &
0.55 & 0.65 \\
&
\multicolumn{2}{>{\centering\arraybackslash}p{(\linewidth - 18\tabcolsep) * \real{0.1866} + 2\tabcolsep}}{%
cLISI} & 1 & 1 & 1 & 1 & 1 & 1 & 1 \\
&
\multicolumn{2}{>{\centering\arraybackslash}p{(\linewidth - 18\tabcolsep) * \real{0.1866} + 2\tabcolsep}}{%
BRAS} & 0.88 & 0.88 & 0.83 & \textbf{0.89} & 0.86 & 0.55 & 0.86 \\
&
\multicolumn{2}{>{\centering\arraybackslash}p{(\linewidth - 18\tabcolsep) * \real{0.1866} + 2\tabcolsep}}{%
iLISI} & \textbf{0.21} & \textbf{0.21} & 0.14 & 0.17 & 0.15 & 0.16 &
0.16 \\
&
\multicolumn{2}{>{\centering\arraybackslash}p{(\linewidth - 18\tabcolsep) * \real{0.1866} + 2\tabcolsep}}{%
KBET} & \textbf{0.44} & 0.48 & 0.17 & 0.31 & 0.18 & 0.18 & 0.31 \\
&
\multicolumn{2}{>{\centering\arraybackslash}p{(\linewidth - 18\tabcolsep) * \real{0.1866} + 2\tabcolsep}}{%
Graph connectivity} & 0.93 & 0.92 & \textbf{0.99} & 0.89 & 0.86 & 0.82 &
0.88 \\
&
\multicolumn{2}{>{\centering\arraybackslash}p{(\linewidth - 18\tabcolsep) * \real{0.1866} + 2\tabcolsep}}{%
PCR comparison} & 0.24 & \textbf{0.51} & 0.15 & 0.02 & 0 & 0 & 0 \\
&
\multicolumn{2}{>{\centering\arraybackslash}p{(\linewidth - 18\tabcolsep) * \real{0.1866} + 2\tabcolsep}}{%
Batch correction} & 0.54 & \textbf{0.6} & 0.45 & 0.46 & 0.41 & 0.34 &
0.44 \\
&
\multicolumn{2}{>{\centering\arraybackslash}p{(\linewidth - 18\tabcolsep) * \real{0.1866} + 2\tabcolsep}}{%
Bio conservation} & 0.76 & 0.75 & \textbf{0.77} & 0.63 & 0.73 & 0.73 &
0.75 \\
&
\multicolumn{2}{>{\centering\arraybackslash}p{(\linewidth - 18\tabcolsep) * \real{0.1866} + 2\tabcolsep}}{%
Total} & 0.67 & \textbf{0.69} & 0.64 & 0.56 & 0.6 & 0.57 & 0.63 \\
\end{longtable}

*NMI, normalized mutual information; ARI, adjusted Rand index; cLISI,
cell-type Local Inverse Simpson's Index; iLISI, integration Local
Inverse Simpson's Index; BRAS, batch removal average silhouette; KBET,
k-nearest neighbor batch effect test; PCR, principal component
regression.
\newpage
\textbf{Table 3. Cross-species benchmarking Nephrobase Cell+ against
existing foundational and dimensionality reduction models in human and
mouse datasets.*}

\begin{longtable}[]{@{}
  >{\centering\arraybackslash}p{(\linewidth - 10\tabcolsep) * \real{0.2092}}
  >{\centering\arraybackslash}p{(\linewidth - 10\tabcolsep) * \real{0.2025}}
  >{\centering\arraybackslash}p{(\linewidth - 10\tabcolsep) * \real{0.2059}}
  >{\centering\arraybackslash}p{(\linewidth - 10\tabcolsep) * \real{0.1429}}
  >{\centering\arraybackslash}p{(\linewidth - 10\tabcolsep) * \real{0.1197}}
  >{\centering\arraybackslash}p{(\linewidth - 10\tabcolsep) * \real{0.1197}}@{}}
\toprule\noalign{}
\begin{minipage}[b]{\linewidth}\centering
Metrics
\end{minipage} & \begin{minipage}[b]{\linewidth}\centering
Nephrobase Cell+ 1B
\end{minipage} & \begin{minipage}[b]{\linewidth}\centering
Nephrobase Cell+ 500M
\end{minipage} & \begin{minipage}[b]{\linewidth}\centering
Geneformer
\end{minipage} & \begin{minipage}[b]{\linewidth}\centering
UCE
\end{minipage} & \begin{minipage}[b]{\linewidth}\centering
scGPT
\end{minipage} \\
\midrule\noalign{}
\endhead
\bottomrule\noalign{}
\endlastfoot
Isolated labels & 0.54 & 0.54 & 0.56 & \textbf{0.64} & 0.56 \\
KMeans NMI & 0.73 & \textbf{0.75} & 0.44 & 0.7 & 0.62 \\
KMeans ARI & 0.57 & \textbf{0.72} & 0.22 & 0.53 & 0.43 \\
Silhouette label & \textbf{0.61} & 0.6 & 0.51 & 0.6 & 0.57 \\
cLISI & 1 & 1 & 0.99 & 1 & 1 \\
BRAS & 0.6 & 0.62 & \textbf{0.69} & 0.38 & 0.64 \\
iLISI & \textbf{0.01} & \textbf{0.01} & 0 & 0 & 0 \\
KBET & \textbf{0.04} & 0.03 & 0.02 & 0 & 0.01 \\
Graph connectivity & \textbf{0.88} & 0.84 & 0.84 & 0.72 & 0.85 \\
PCR comparison & \textbf{0.94} & 0.94 & 0.61 & 0.04 & 0.8 \\
Batch correction & \textbf{0.49} & 0.49 & 0.43 & 0.23 & 0.46 \\
Bio conservation & 0.69 & \textbf{0.72} & 0.54 & 0.69 & 0.64 \\
Total & 0.61 & \textbf{0.63} & 0.5 & 0.51 & 0.57 \\
\end{longtable}

*NMI, normalized mutual information; ARI, adjusted Rand index; cLISI,
cell-type Local Inverse Simpson's Index; iLISI, integration Local
Inverse Simpson's Index; BRAS, batch removal average silhouette; KBET,
k-nearest neighbor batch effect test; PCR, principal component
regression.

\end{document}